\documentclass[aps,prb,twocolumn, nofootinbib, superscriptaddress,prb,floatfix]{revtex4-2}
\usepackage{amsmath,amsthm,amsfonts,amssymb,amscd,bbold}
\usepackage[T1]{fontenc}
\usepackage{bm}
\usepackage{mathrsfs} 
\usepackage{dcolumn}  
\usepackage{physics}
\usepackage{graphicx}
\usepackage{epstopdf}
\usepackage{float}
\usepackage{pgf}
\usepackage{tabularx}
\usepackage{subfigure}
\usepackage{float}
\usepackage{hyperref}

\newcommand{\dsp}{\displaystyle}

\begin{document}
\title{Electrical operation of hole spin qubits in planar MOS silicon quantum dots}

\author{Zhanning Wang}
\affiliation{School of Physics, The University of New South Wales, Sydney NSW 2052, Australia}

\author{Abhikbrata Sarkar}
\affiliation{School of Physics, The University of New South Wales, Sydney NSW 2052, Australia}
\affiliation{Australian Research Council Centre of Excellence in Future Low-Energy Electronics Technologies, The University of New South Wales, Sydney NSW 2052, Australia}

\author{S. D. Liles}
\affiliation{School of Physics, The University of New South Wales, Sydney NSW 2052, Australia}

\author{Andre Saraiva}
\affiliation{School of Electrical Engineering and Telecommunications, The University of New South Wales, Sydney NSW 2052, Australia}

\author{A. S. Dzurak}
\affiliation{School of Electrical Engineering and Telecommunications, The University of New South Wales, Sydney NSW 2052, Australia}

\author{A. R. Hamilton}
\affiliation{School of Physics, The University of New South Wales, Sydney NSW 2052, Australia}
\affiliation{Australian Research Council Centre of Excellence in Future Low-Energy Electronics Technologies, The University of New South Wales, Sydney NSW 2052, Australia}

\author{Dimitrie Culcer}
\affiliation{School of Physics, The University of New South Wales, Sydney NSW 2052, Australia}
\affiliation{Australian Research Council Centre of Excellence in Future Low-Energy Electronics Technologies, The University of New South Wales, Sydney NSW 2052, Australia}

\date{\today}
\begin{abstract}
Silicon hole quantum dots have been the subject of considerable attention thanks to their strong spin-orbit coupling enabling electrical control, a feature that has been demonstrated in recent experiments combined with the prospects for scalable fabrication in CMOS foundries. The physics of silicon holes is qualitatively different from germanium holes and requires a separate theoretical description, since many aspects differ substantially: the effective masses, cubic symmetry terms, spin-orbit energy scales, magnetic field response, and the role of the split-off band and strain. In this work, we theoretically study the electrical control and coherence properties of silicon hole dots with different magnetic field orientations, using a combined analytical and numerical approach. We discuss possible experimental configurations required to obtain a sweet spot in the qubit Larmor frequency, to optimize the electric dipole spin resonance (EDSR) Rabi time, the phonon relaxation time, and the dephasing due to random telegraph noise. Our main findings are: (i) The in-plane $g$-factor is strongly influenced by the presence of the split-off band, as well as by any shear strain that is typically present in the sample. The $g$-factor is a non-monotonic function of the top gate electric field, in agreement with recent experiments. This enables coherence sweet spots at specific values of the top gate field and specific magnetic field orientations. (ii) Even a small ellipticity (aspect ratios $\sim 1.2$) causes significant anisotropy in the in-plane $g$-factor, which can vary by $50\% - 100\%$ as the magnetic field is rotated in the plane. This is again consistent with experimental observations. (iii) EDSR Rabi frequencies are comparable to Ge, and the ratio between the relaxation time and the EDSR Rabi time $\sim 10^5$. For an out-of-plane magnetic field the EDSR Rabi frequency is anisotropic with respect to the orientation of the driving electric field, varying by $\approx 20\%$ as the driving field is rotated in the plane. Our work aims to stimulate experiments by providing guidelines on optimizing configurations and geometries to achieve robust, fast and long-lived hole spin qubits in silicon.
\end{abstract}
\maketitle

\section{Introduction}
\label{Section 1 - Introduction}

Silicon quantum devices have emerged as an ideal platform for scalable quantum computation, with remarkable advancements both theoretically and experimentally in recent years \cite{Loss1998,kane1998,Schofield2001,Ladd2002,Hanson2007,Morton2011,Zwanenburg2013,Kalra2014,Hill2015,Salfi2016,Salfi20162,Veldhorst2017,Hutin2018,Niquet2018,Yoneda2018,Takashi2018,Abadillo2018,Keith2019,Hendrickx2020,Hendrickx20202,Kodera2020,Laucht2021,Burkard2021,Kobayashi2021,Zalba2021,Chatterjee2021,Aggarwal2021,Fang2022,Petta2022,Sarkar2022,Lodari2022,Borsoi2022,Ciocoiu2022,Krauth2022,Becher2022}. Silicon devices offer several advantages, including weak hyperfine interaction with the possibility of isotopic purification to eliminate the hyperfine coupling altogether \cite{Tyryshkin2012,Chekhovich2013,Prechtel2016,Yoneda2018,Chesi2020,Bosco20212,Noiri2022}, absence of piezoelectric phonons \cite{Tahan2014,Venitucci2020}, and mature silicon micro-fabrication technology \cite{Fuhrer2009,Pla2012,Maune2012,Veldhorst2015,Jehl2016,Amitonov2018,Watson2018,He2019,Tanttu2019,Crippa2019,Fricke2021}, making them competitive candidates to realize industrial-level scalable quantum computing architectures. Over the past few decades, numerous design proposals for qubits utilizing silicon quantum devices have been actively investigated, including the singlet-triplet transition qubit \cite{Maune2012}, single electron spin qubit \cite{Pla2012,Kawakami2014,Takeda2016}, and acceptor or donor spin qubit \cite{Mahapatra2013,Kalra2014,Salfi2014,Usman2015,Salfi2016,He2019}. Among the various platforms, silicon hole spin qubits exhibit additional desirable properties \cite{Ruess2004,Roddaro2008,Steele2009,Fang2009,Simon2009,Dery2011,Kloeffel20132,Tahan2014,Higginbotham2014,Voisin2016,Tanamoto2017,Kloeffel2018,Yoneda2020,Milivojevic2021,Froning2021,Duan2021,Malkoc2022,Shalak2023}. Firstly, hole systems possess strong spin-orbit coupling \cite{Winkler2000,Winkler2002,Danneau2006,Culcer2006,Charkraborty2009,Kloeffel2011,Kloeffel2013,Chesi2014,Durnev2014,Moriya2014,Bihlmayer2015,Miserev2017,Miserev20172,Marcellina2018,Tanttu2019,Carlos2022,Lidal2023}, which enables pure electrical manipulation of spin states via EDSR \cite{Golovach2006,Bulaev2007,Rashba2008,Rashba2012}, while the hole spin-3/2 is responsible for physics with no counterpart in electron systems \cite{Winkler2003,Culcer2006,Winkler2008, Hong2018,Abadillo2018,Cullen2021,Sina2023,Farokhnezhad2023}. Secondly, the absence of valley degeneracy avoids complications associated with the increase in Hilbert space that occurs for electrons \cite{McGuire2007, Chutia2007, Cywinski2010,Culcer2010, Friesen2010,Hao2014,Salfi2014,Boross2016, Niquet2018, Ferdous2018, Voisin2020, Spence2023,Voisin2022}. Thirdly, whereas the hyperfine interaction is a strong decoherence source in other materials such as III-V group semiconductors \cite{Itoh1993,Khaetskii2002,Coish2004,Petta2005,Laird2007,Fischer2008,Baugh2009, Fischer2010,Chekhovich2011,Assali2011,Chekhovich2013,Kloeffel2013,Prechtel2016,Wang2016}, silicon can be isotopically purified \cite{Itoh2003,Tyryshkin2012,Kawakami2014,Takeda2016,Veldhorst2017,Hutin2018,Piot2022}. Recent years have witnessed key experiments on silicon hole qubits, including successful demonstrations on industrial standard complementary metal-oxide-semiconductor (CMOS) technologies \cite{Horibe2015,Voisin2016,Veldhorst2017,Tanamoto2017,Zumbuhl2018,Amitonov2018,Hutin2018,Crippa2018,Ruoyu2020,Wei2020,Spence20222,Ik2023}, control of the number of holes and shell filling \cite{Fang2009,Liles2018}, $g$-tensor manipulation in both nanowire and quantum dot systems \cite{Voisin2016,Crippa2018,Wei2020,Liles2021}, qubit operation at 25\,K in the few-hole regime \cite{Shimatani2020}, single qubit operation above 4\,K \cite{Camenzind2022}, long coherence time up to 10\,ms in Si:B acceptors \cite{Kobayashi2021}, dispersive readout \cite{Tanamoto2017,Pakkiam2018,Crippa2019, Duan2021,Ezzouch2021,Russell2022}, Pauli spin blockade \cite{Li2015,Corna2016}, coupling between photons and hole spins \cite{Yu2022}, and the demonstration of the coupling between two hole qubits via anisotropic exchange \cite{Geyer2022}.

In parallel with developments in silicon, considerable attention has been devoted to hole spin qubits in germanium \cite{Pillarisetty2011,Dobbie2012,Lada2018,Mizokuchi2018,Hendrickx2018,Sammak2019,Lodari2019,Hardy2019,Lawrie2020,Gao2020,Xu2020,Hendrickx20202,Terrazos2021,Scappucci2021,Jirovec2021,Wang2021,Hendrickx2021,Mutter2021,Bosco2021,Ungerer2022,Liu2022,Deprez2022,Martinez2022,Adelsberger20222,Deprez2022}. This includes spin state measurement and readout \cite{Roddaro2008,Fang2009,Zarassi2017,Li2017,Katsaros2020,Froning20212}, electrical control of spin states \cite{Pribiag2013}, $g$-tensor manipulation \cite{Ares2013,Brauns2016,Corna2016,Watzinger2016,Lu2017,Watzinger2018,Mizokuchi2018,Zhang2021}, coupling to a superconducting microwave resonator \cite{Li2018}, fast EDSR Rabi oscillations up to 540\,MHz \cite{Ke2022} and relaxation times of up to 32\,ms \cite{Lawrie2020,Lawrie20202}. Nevertheless, the understanding gained from the study of germanium hole qubits cannot be directly translated to silicon. In silicon, the split-off energy is very small (44\,meV) compared with gallium arsenide (340\,meV) and germanium (296\,meV) \cite{Phillips1962,Cardona1966}, necessitating the use of a full six-band Luttinger-Kohn model\cite{Luttinger1955,Baldereschi1971,Baldereschi1973,Chuang1995,Cheng2005,Schilfgaarde2005,Bulaev20052,Korkusinski2008, Shen2010, Marcellina2017}. Additionally, the larger effective mass of holes in silicon requires smaller dots to achieve the same orbital confinement energy splitting. The strength of the spin-orbit coupling in silicon is weaker than in germanium, while the cubic symmetry parameters are very strong and cannot be accounted for perturbatively. Moreover, previous studies have shown that, for silicon two-dimensional hole gases at experimentally relevant densities, the Schrieffer-Wolff transformation cannot be used to reduce the three-dimensional Hamiltonian to an effective two-dimensional one since the criteria for the applicability of the Schrieffer-Wolff transformation are not satisfied \cite{Marcellina2017, Akhgar2019}. Furthermore, the orbital magnetic field terms are not captured by such a Schrieffer-Wolff transformation, which will play an important role when the magnetic field is applied perpendicular to the gate electric field. Finally, strain effects in silicon are different from those in other materials, with axial and shear strains strongly affecting both spin dynamics and the in-plane $g$-factor in planar quantum dots. In particular, spatial strain gradients caused by thermal contraction of the gate electrodes have a very large effect in silicon due to the thin gate oxide \cite{Liles2021}.

Theoretical studies on silicon hole qubits have also advanced rapidly in line with experimental progress. The physics of spin-orbit coupling in silicon hole systems have been investigated with the aim of realizing fully electrically operated spin qubits \cite{Marcellina2017, Kloeffel2018}. Studies have examined device geometry, dot orientation, and strain in the silicon quantum dot to optimize the quality of the EDSR Rabi frequency and qubit Larmor frequency. These studies have identified optimal operation points and the possibility of achieving fast EDSR Rabi oscillations even with small spin-orbit coupling in silicon hole systems \cite{Niquet2019,Milivojevic2021,Moratis2021,Bosco20213,Froning2021,Qvist2022,Adelsberger2022,Bosco2022,Platero2022}. Decoherence due to hyperfine interactions and dephasing due to charge noise have also been studied, identifying experimental configurations for ultra-fast and highly coherent silicon hole spin qubits \cite{Chesi2020, Venitucci2020, Bosco20212, Malkoc2022,Shalak2023}. However, despite significant progress in both experiment and theory, a critical question remains unanswered: \textit{When is it possible to minimize unwanted decoherence effects without reducing the efficiency of EDSR?}

\begin{figure}[tbp!]
\includegraphics[width=0.48\textwidth]{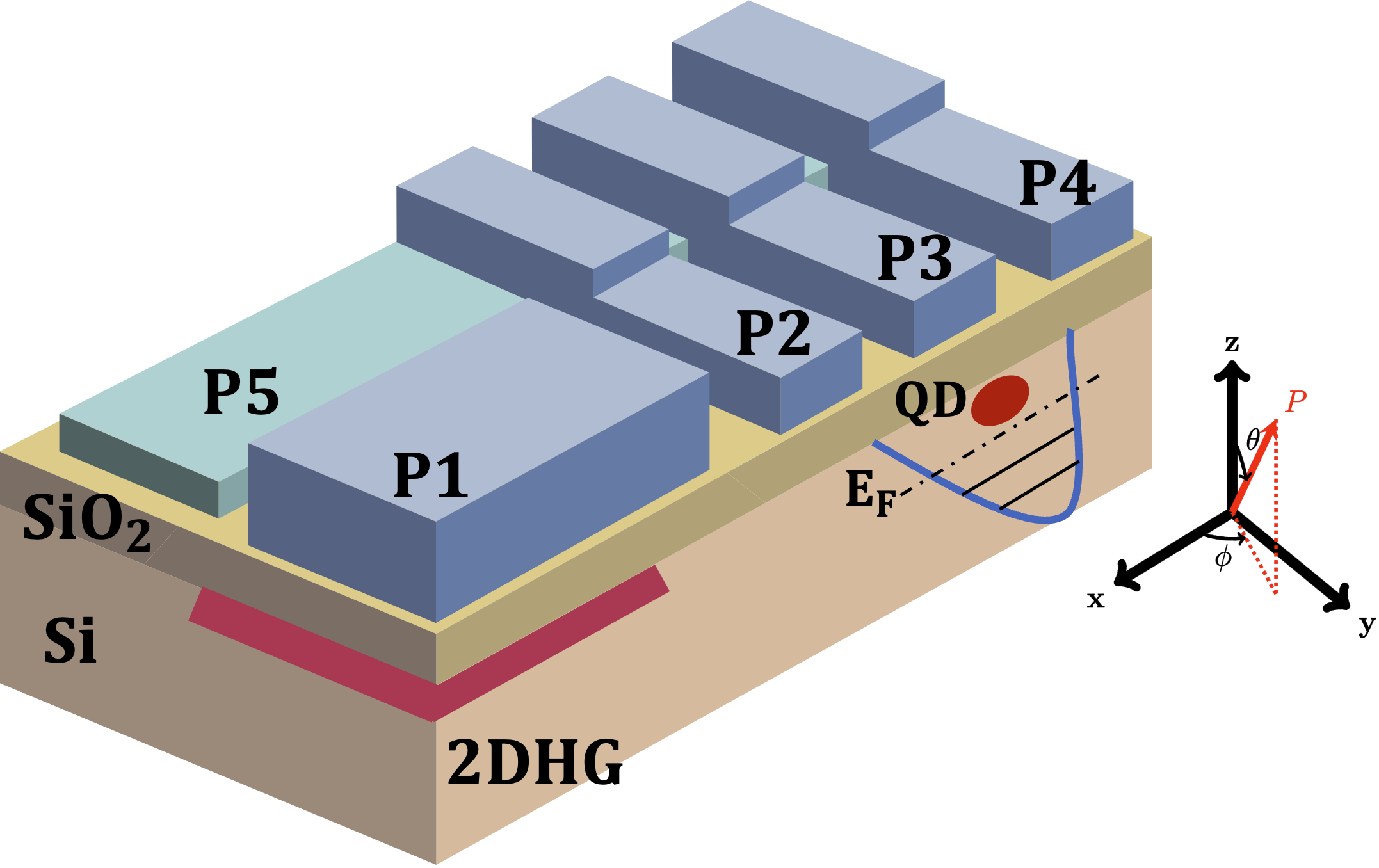}
\caption{\textbf{Schematic planar silicon quantum dot}. In this specific design, we focus on a single hole quantum dot in the silicon layer. By applying a gate electric field $F_z$ via gate P1, holes accumulate in silicon and are confined vertically against the silicon oxide (indicated at the location of the two-dimensional hole gas). The single quantum dot is formed using gates P2-P5. The gates P2 and P4 provide confinement in the $\hat{x}$-direction, while gate P5 provides confinement in the y direction. P3 is used as the top gate of the quantum dot, accumulating a single hole in the potential well beneath. The resulting potential is indicated schematically below the gates.
}
\label{fig: Silicon Qubit Prototype}
\end{figure}

In this paper, we focus on electrically-driven single hole spin qubits in planar silicon quantum dots and describe qubit dynamics in both perpendicular and in-plane magnetic fields. We adopt a hybrid analytical and computational approach which enables us to treat quantum dots with arbitrary confinement in a magnetic field of arbitrary orientation. For a perpendicular magnetic field we show that coherence sweet spots exist at certain values of the top gate field, which reflect the coupling of heavy- and light-hole states by the gate electric field. The EDSR Rabi frequency exhibits a maximum as a function of the top gate field, as does the relaxation rate. The large Rabi ratios (the ratio between the phonon relaxation time and the EDSR Rabi time) can be achieved, in excess of $10^6$ at very small in-plane driving electric fields of 1\,kV/m. For an in-plane magnetic field, we demonstrate that the qubit Zeeman splitting exhibits a large modulation as a function of the top gate electric field. Although extrema in the qubit Zeeman splitting exist as a function of the top gate field, these do not protect against charge noise, and one cannot identify coherence sweet spots, since the qubit is exposed to all three components of the noise electric field. At the same time, we find that the EDSR Rabi frequency reaches a maximum of about 100\,MHz, with a minimum relaxation time of 1\,ms, yielding a Rabi ratio of approximately $10^5$ for an in-plane driving electric field of 1\,kV/m. Importantly, the $g$-factor of elliptical dots is strongly anisotropic, with a very small aspect ratio (1.2) yielding a factor of 0.7-1.6 variation as the magnetic field is rotated in the plane. This is consistent with recent experimental observations \cite{Liles2021}. Finally, we compare the properties and fabrication technologies of silicon and germanium and demonstrate that shear strain and axial strain are key factors leading to a large modulation of in-plane $g$-factors and the large Rabi ratio. The EDSR Rabi frequencies for a given in-plane driving electric field are comparable in the two materials, which may reflect the fact that, while Ge has stronger spin-orbit coupling, Si has larger cubic symmetry terms $\propto (\gamma_3 - \gamma_2)$, which enhance the effective spin-orbit coupling experienced by planar dots. Whereas we use characteristic values of the strain tensor components extracted from experiment, further investigation is needed to understand the role of strain and of the strain distribution throughout the sample.

The manuscript is organized as follows. In Section\,\ref{Section 1 Model and Methodology}, we introduce the Hamiltonian for the silicon hole quantum dot with an arbitrary magnetic field orientation, discuss the diagonalization technique, and outline the methodology used to determine the EDSR Rabi frequency, relaxation time due to phonons, and dephasing time due to random telegraph noise. In Section\,\ref{Result and Discussion}, we present the results in the presence of a perpendicular magnetic field as well as an in-plane magnetic field. We discuss the effect of ellipticity of the quantum dot and $g$-factor anisotropy in line with experimental observations. In Section \ref{Comparison between Germanium and Silicon}, we compare the properties of silicon hole qubits with germanium hole qubits from the perspective of material parameters and fabrication details. We end with a summary and conclusions.

\section{Model and Methodology} 
\label{Section 1 Model and Methodology}

In this section, we elucidate the properties of a single silicon hole spin qubit by introducing the model device and relevant experimental parameters. Furthermore, we provide a detailed discussion on the physical origin of the strain Hamiltonian, confinement Hamiltonian, and Zeeman Hamiltonian, respectively, in the context of the envelope function approximation Hamiltonian $H$. We also introduce the details of the numerical diagonalization used to obtain the relevant energy levels and wave-functions of the system. Then, we present the formalism used to estimate the EDSR Rabi frequency, relaxation time, and dephasing time.

A schematic diagram of a possible realization of a silicon hole spin qubit is described in Fig.\,\ref{fig: Silicon Qubit Prototype}. The gate electric field is applied along the $\hat{z}$-direction, denoted by $\bm{F}\,=\,(0,0,F_z)$. Our model is designed to describe a generic magnetic field $\bm{B}\,=\,(B_x, B_y, B_z)$ as illustrated in Fig.\,\ref{fig: Silicon Qubit Prototype}, using the vector potential $\bm{A}\,=\,-(B_z y, B_x z, B_y x)$. We consider magnetic fields either in the $xy$-plane or parallel to the $\hat{z}$-direction (perpendicular to the qubit plane); we do not consider magnetic fields tilted out of the plane in this work.

\subsection{Diagonalization of silicon hole spin qubit Hamiltonian}
\label{Diagonalization of silicon hole spin qubit Hamiltonian}

The total Hamiltonian for a single silicon hole quantum dot qubit is given by $H=H_{\text{LKBP}} + H_{\text{conf}} + H_{\text{gate}} + H_{\text{Zeeman}}$. The perpendicular electric confinement potential is represented by $H_{\text{gate}}\,=\,e F_z z$ for z $\in [-L/2, L/2]$, where $L$ denotes the width of the quantum well in the $\hat{z}$-direction. This gate field induces structural inversion asymmetry (SIA) in the silicon hole system, thereby leading to a Rashba spin-orbit coupling. Moreover, the symmetry of the diamond lattice ensures there is no Dresselhaus-type spin-orbit coupling \cite{Dresselhaus1955, Ohkawa1974, Bychkov1984,Governale2002,Meier2007,Morrison2014,Nitta2015}. Although interface misalignment may induce Dresselhaus-type spin-orbit coupling, it is expected to be negligible for the purposes of this paper \cite{Marcellina2017}, and is therefore not considered. In-plane confinement is modeled by a two-dimensional harmonic oscillator potential $H_{\text{conf}}\,=\,\hbar^2 x^2/(2 m^* a_x^4) + \hbar^2 y^2/(2 m^* a_x^4)$, where $m^*$ is the in-plane effective mass, and $a_x$ and $a_y$ are the two axes of an elliptical dot.

The Zeeman Hamiltonian can be written as $H_{\text{Zeeman}}\,=\,-2 \kappa_1 \mu_B \bm{J}\cdot\bm{B} - 2 \kappa_2 \mu_B \bm{J}_3\cdot\bm{B}$, where $\bm{J}\,=\,(J_x, J_y, J_z)$ represents the angular momentum matrices for the direct sum of the spin-3/2 and spin-1/2 systems. The explicit matrix form of expressions for $\bm{J}$ can be found in the supplementary material. Additionally, in the an-isotropic term in the Zeeman Hamiltonian, we have $\bm{J}_3\,=\,(J_x^3, J_y^3, J_z^3)$. $\mu_B$ is the Bohr magneton, and $\kappa_1\,=\,-0.42$, $\kappa_2\,=\,0.01$ for silicon.

The quantum dot studied in this paper is produced by confining a two-dimensional hole gas in a metal-oxide-semiconductor structure grown along the $\hat{z}$ $\parallel$[001] direction. The strong heavy hole - light hole splitting results in angular momentum quantization perpendicular to the two-dimensional plane. The heavy-hole states are characterized by a $\hat{z}$-component of the angular momentum $\ket{J=3/2,M=\pm 3/2}$, the light-hole states are characterized by $\ket{J=3/2,M=\pm 1/2}$, while the split-off valence band has $\ket{J=1/2, M=\pm 1/2}$. We orient the wave vectors $k_x, k_y, k_z$ along [100], [010], and [001], respectively. The valence band and the effect of strain can be described by the Luttinger-Kohn-Bir-Pikus (LKBP) Hamiltonian in the basis of total angular momentum eigenstates $\{\left|\frac{3}{2}, \frac{3}{2}\right\rangle,\left|\frac{3}{2}, \frac{1}{2}\right\rangle,\left|\frac{3}{2},-\frac{1}{2}\right\rangle,\left|\frac{3}{2},-\frac{3}{2}\right\rangle,\left|\frac{1}{2}, \frac{1}{2}\right\rangle,\left|\frac{1}{2},-\frac{1}{2}\right\rangle\}$:
\begin{small}
\begin{eqnarray}\label{Eq - HLKBP}
\hspace*{-1cm}  &&H_{\text{LKBP}}=\nonumber\\
\hspace*{-1cm}   &&\begin{bmatrix}
P+Q & 0 & -S & R & -\displaystyle\frac{1}{\sqrt{2}} S & \displaystyle\sqrt{2} R \\
0 & P+Q & R^* & S^* & -\sqrt{2} R^* & -\displaystyle\frac{1}{\sqrt{2}} S^* \\
-S^* & R & P-Q & 0 & -\sqrt{2} Q & \displaystyle\sqrt{\frac{3}{2}} S \\
R^* & S & 0 & P-Q & \displaystyle\sqrt{\frac{3}{2}} S^* & \sqrt{2} Q \\
-\displaystyle\frac{1}{\sqrt{2}} S^* & -\sqrt{2} R & -\sqrt{2} Q^* & \displaystyle\sqrt{\frac{3}{2}} S & P+\Delta_0 & 0 \\
\sqrt{2} R^* & -\displaystyle\frac{1}{\sqrt{2}} S & \displaystyle\sqrt{\frac{3}{2}} S^* & \sqrt{2} Q^* & 0 & P+\Delta_0
\end{bmatrix}
\end{eqnarray}
\end{small}
where $P\,=\,P_k + P_\varepsilon, Q\,=\,Q_k + Q_\varepsilon, R\,=\,R_k + R_\varepsilon, S\,=\,S_k + S_\varepsilon$, $\Delta_0\,=\,44$\,\text{meV} is the energy splitting between the heavy-hole band and the split-off band, which is small enough to lead to the necessity of extending calculations to the six-band LKBP Hamiltonian (in germanium this is usually not necessary). Terms with subscripts $k$ are matrix elements from the Lutitinger-Kohn (LK) Hamiltonian \cite{Luttinger1955}: $P_k\,=\,\frac{\hbar^2}{2 m_0}\gamma_1\left(k_x^2+k_y^2+k_z^2\right)$, $Q_k\,=\,-\frac{\hbar^2}{2 m_0} \gamma_2\left(2 k_z^2-k_x^2-k_y^2\right)$ appears in the diagonal elements; while $R_k= \frac{\sqrt{3}\hbar^2}{2 m_0}[-\gamma_2\left(k_x^2-k_y^2\right)$ $+2 i \gamma_3 k_x k_y]$, $S_k\,=\, \frac{\sqrt{3}\hbar^2}{m_0} \gamma_3\left(k_x-i k_y\right) k_z$ appears in the off-diagonal elements which couple the heavy-hole bands to light-hole bands and split-off bands. Here $\gamma_1\,=\,4.285, \gamma_2\,=\,0.339, \gamma_3\,=\,1.446$ are the Luttinger parameters for silicon \cite{osten1987,Soline2004}, $m_0$ is the bare electron mass and $\hbar$ is the Planck constant. Terms with subscripts $\varepsilon$ are matrix elements from the Bir-Pikus (BP) Hamiltonian: $P_{\varepsilon}\,=\,-a_v\left(\varepsilon_{xx}+\varepsilon_{yy}+\varepsilon_{zz}\right)$, $Q_{\varepsilon}\,=\,-\frac{b_v}{2}\left(\varepsilon_{xx}+\varepsilon_{yy}-2 \varepsilon_{zz}\right)$, $R_{\varepsilon}\,=\,\frac{\sqrt{3}}{2} b_v\left(\varepsilon_{xx}-\varepsilon_{yy}\right)-i d_v \varepsilon_{xy}$, $S_{\varepsilon}\,=\,-d_v\left(\varepsilon_{xz}-i \varepsilon_{yz}\right)$. The material parameter $a_v\,=\,2.38\,$eV is the hydro-static deformation potential constant, $b_v\,=\,-2.10\,$eV is the uni-axial deformation potential constant, $d_v\,=\,-4.85\,$eV is the shear deformation potential constant \cite{Wortman1965,Hopcroft2010,Cedric2023}. The strain $\varepsilon_{i,j} $ where $i,j \in \{x,y,z\}$ is determined by experimental configurations and fabrication processes. 

In a quantum dot placed in a magnetic field, the momentum operators in the Luttinger-Kohn Hamiltonian are modified by the gauge potentials. The new canonical conjugate momentum operators are given by $\bm{p} + e \bm{A}$. To numerically diagonalize the total Hamiltonian $H$, the wave functions we used are as follows:
\begin{equation}\label{Eq - total wave-function}
    \Psi_{n_x, n_y, n_z, i}(x,y,z)\,=\,\phi_{n_z}(z)\phi_{n_x}(x)\phi_{n_y}(y) \ket{\chi_i}
\end{equation}
where $n_x, n_y, n_z$ are the level numbers of the spatial wave functions and $\ket{\chi_i}$ is the i-th spinor.

The selection of wave functions depends on the shape of the confinement potentials, which necessitates the self-consistent solution of Poisson and Schrödinger equations to account for the density-dependent properties of a device. Previous studies have demonstrated the efficiency of the variational approach in gallium arsenide and germanium \cite{Marcellina2017}. However, the numerical generalization of the variational method becomes challenging as the number of energy levels increases. Our approach is to determine a set of complete wave functions in all directions, using a sufficient number of energy levels to include the geometry of the quantum confinements.

For $\hat{z}$-direction, where our focus is on a triangular quantum well $eF_zz$, we select sinusoidal wave functions derived from an infinite square well positioned symmetrically between $z \in [-L/2, L/2]$. Incorporating the boundary conditions, these orthonormal complete set of wave functions are
\begin{equation}\label{Eq - z wave functions 1}
\phi_{n_z}(z)\,=\,\sqrt{\frac{2}{L}} \cos \left(\frac{n_z \pi z}{L}\right) \quad \text { for } n_z=1,3,5, \dots
\end{equation}
\begin{equation}\label{Eq - z wave functions 2}
\phi_{n_z}(z)\,=\,\sqrt{\frac{2}{L}} \sin \left(\frac{n_z \pi z}{L}\right) \quad \text { for } n_z=2,4,6, \dots
\end{equation}
The in-plane wave functions we use are eigenstates of the two-dimensional harmonic oscillator:
\begin{equation}\label{Eq - x wave functions 3}
    \phi_{n_x}(x) =\frac{1}{\sqrt{2^n n !}}\sqrt[4]{\frac{1}{\pi}} \frac{1}{\sqrt{a_x}} \exp \left(-\frac{x^2}{2 a_x^2}\right) H_{n_x}\left(\frac{x}{a_x}\right),
\end{equation}
\begin{equation}\label{Eq - y wave functions 4}
    \phi_{n_y}(y) =\frac{1}{\sqrt{2^n n !}}\sqrt[4]{\frac{1}{\pi}} \frac{1}{\sqrt{a_y}} \exp \left(-\frac{y^2}{2 a_y^2}\right) H_{n_y}\left(\frac{y}{a_y}\right),
\end{equation}
where the confinement frequency can be expressed as $\omega_{x,y}\,=\,\hbar/(m^* a^2_{x,y})$, $H_{n}$ represents Hermitian polynomials. To obtain high accuracy in numeric, we adopt 8 levels of Eq.\,\ref{Eq - z wave functions 1}, 14 levels of Eq.\ref{Eq - x wave functions 3}, and 14 levels of \ref{Eq - y wave functions 4}. We can then diagonalize the Hamiltonian $H$ for specific device geometries, strains, electric fields, and magnetic fields. The ground state of the qubit Hamiltonian will be denoted by $\ket{\mathbb{0}}$ with energy $E_{\mathbb{0}}$, the first excited state will be denoted by $\ket{\mathbb{1}}$ with energy $E_{\mathbb{1}}$ and the qubit Zeeman splitting will be defined by $\Delta E_z\,=\,E_{\mathbb{1}}-E_{\mathbb{0}}$. Consequently, the total Hamiltonian $H$ will be diagonalized in the basis $\{\ket{\mathbb{0}}, \ket{\mathbb{1}}, \ket{\mathbb{2}},...,\ket{\mathbb{N}} \}$.

\subsection{EDSR frequency and phonon relaxation time}
\label{EDSR frequency and phonon relaxation time}

Electric dipole spin resonance (EDSR) methods are widely used to coherently drive transitions between spin states in silicon hole spin qubits. Within the ground state orbital, the spin of a hole qubit can be rotated by an alternating microwave signal (an alternating in-plane electric field), the frequency of this microwave signal should be matched with the qubit Zeeman splitting $\Delta E_z$. Therefore, the EDSR frequency can be calculated by evaluating the transition matrix element
\begin{equation}\label{Eq - EDSR}
    f\,=\,\frac{e}{h} \norm{\mel**{\mathbb{1}}{\bm{r} \cdot \bm{E}_{\text{AC}}}{\mathbb{0}}}
\end{equation}
In our case, the in-plane alternating driving electric field $E_{\text{AC}}$ is set to be 1\,kV/m. 

The phonon relaxation time can be calculated using the method detailed in numerous studies \cite{Khaetskii2000, Woods2002, Schilfgaarde2005, Cheng2005, Tahan2005, Bulaev2005, Bulaev20052, Stano2006, Simon2009, Shen2010, Hu2012, Venitucci2020}. Unlike III-V semiconductors, silicon and germanium do not have piezoelectric phonons due to their non-polar nature \cite{Tahan2014, Venitucci2020}. We assume that the silicon hole spin qubit is coupled with a thermal bath of bulk acoustic phonons along the polarization direction $\alpha \in { \ell,t_1,t_2}$ (one longitudinal direction and two transverse directions), with phonon wave vectors denoted by $\bm{q}$. The energy of the acoustic phonons is $\hbar \omega_{\alpha,\bm{q}}$. To calculate the phonon relaxation time, we consider the hole-phonon interaction Hamiltonian $H^\alpha_{\text{hp}}=\sum_{i,j} D_{i,j} \varepsilon^\alpha_{i,j}(\bm{r})$ for $i,j \in \{x,y,z\}$, where $D_{i,j}$ are deformation potential matrices. The detailed matrix elements can be found in the supplemental material. The local strain $\varepsilon^\alpha_{i,j}(\bm{r})$ has the form
\begin{equation}\label{Eq - strain SQ}
    \varepsilon^\alpha_{i,j}\,=\,\frac{i}{2} q \sqrt{\frac{\hbar}{2 \rho V_c \omega_{\alpha,\bm{q}}}} \epsilon^\alpha_{i,j} \pqty{ e^{-i \bm{q} \cdot \bm{r} } \hat{a}^\dagger_{\alpha,\bm{q}} + e^{i \bm{q} \cdot \bm{r} } \hat{a}_{\alpha,\bm{q}} }
\end{equation}
where 
\begin{equation}\label{Eq - Dimensionless strain}
    \epsilon^\alpha_{i,j}\,=\, \hat{e}_i^\alpha \frac{q_j}{q} + \hat{e}_j^\alpha \frac{q_i}{q}
\end{equation}
is a symmetric $3\times3$ matrix, $\hat{e}$ is the unit vector in the direction of phonon propagation. The transition rate between the first excited state $E_{\mathbb{1}}$ and the ground state $E_{\mathbb{0}}$ due to spontaneous phonon emission can be obtained from Fermi's golden rule:
\begin{widetext}
\begin{align}\label{Eq - Relaxation rate}
    \Gamma^\alpha\,=\,\frac{2\pi}{\hbar} \int_0^\infty \int_0^{2\pi} \int_0^\pi \norm{ \mel**{\mathbb{0},N_q+1}{H^\alpha_{\text{hp}}}{\mathbb{1},N_q} }^2 q^2 \delta\bqty{ \pqty{ \frac{ E_{\mathbb{1}} - E_{\mathbb{0}} }{ \hbar v_\alpha } }- q } \sin\theta \frac{V_c}{(2\pi)^3} \dd{\theta} \dd{\phi} \dd{q}
\end{align}
\end{widetext}
where $V_c/(2\pi)^3$ is the reciprocal space density of states, $N^\alpha_q$ is the phonon occupation number following Bose-Einstein statistics: $N^\alpha_q=\bqty{\exp\pqty{\hbar \omega^\alpha_{\bm{q}}/\beta }-1}^{-1}$, $\beta\,=\,k_B T$, where T\,=\,0.1 K, and $k_B$ is the Boltzmann constant. The phonon propagation velocity is different for different polarization directions, with $v_\ell\,=\,9000\,$m/s for the longitudinal direction, $v_{t_1}\,=\,v_{t_2}\,=\,5400\,$m/s for the other two transversal directions \cite{Wortman1965,Hao2000,Hopcroft2010}.

\begin{figure}[tbp!]
\includegraphics[width=0.48\textwidth]{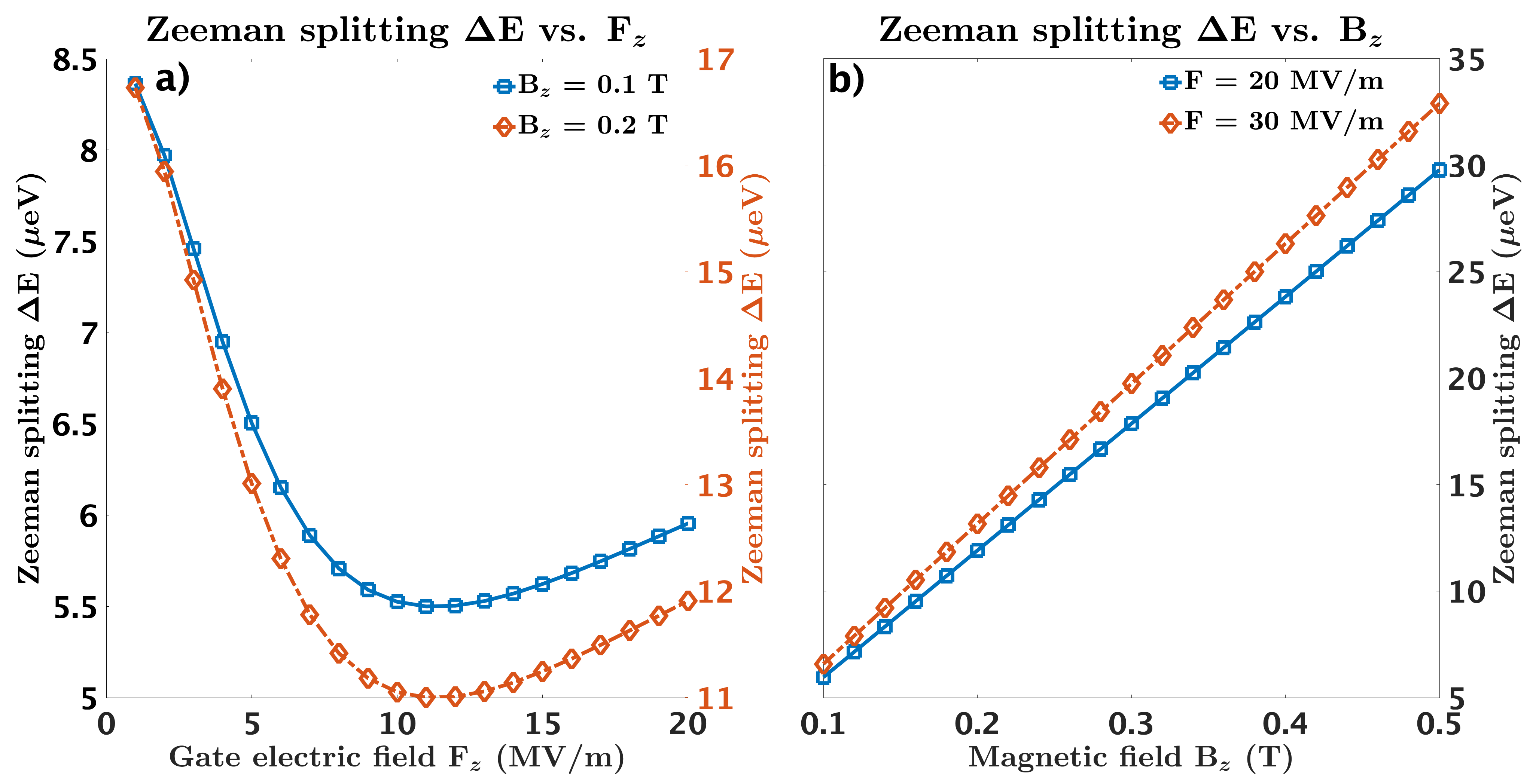}
\caption{\textbf{Qubit Zeeman splitting for an out-of-plane magnetic field $\bm{\mathrm{B}_z}$}. 
a) The qubit Zeeman splitting is plotted as a function of the gate electric field for two different out-of-plane magnetic field strengths, B${}_z$\,=\,1\,T (solid line with square markers) and B${}_z$\,=\,0.8\,T (dashed line with diamond markers). A flat local minimum of the qubit Zeeman splitting is observed as a function of the gate electric field. 
b) The qubit Zeeman splitting is plotted as a function of the out-of-plane magnetic field B${}_z$\,=\,1\,T for two different top gate field strengths, F${}_z$\,=\,20\,MV/m (solid line with square markers) and F${}_z$\,=\,30\,MV/m (dashed line with diamond markers). The parameters used to generate all the figures in this paper are provided in Table\,\ref{TB1 - Silicon VS Germanium}
}
\label{fig: QZS OP All}
\end{figure}

\subsection{Random telegraph noise coherence time}
\label{Random telegraph noise coherence time}
In a silicon hole quantum dot system, the spin-orbit coupling induced by the top gate field exposes the hole spin qubit to charge noise, primarily from charge defects. The charge defects can lead to fluctuation in the qubit energy spectrum, resulting in qubit dephasing \cite{Culcer2009, Ramon2010, Culcer2013, Bermeister2014, Roszak2019,Lodari2021,Malkoc2022}. In our model, we particularly focus on two key sources of charge defects-induced dephasing: screened single charge defects and dipole charge defects. 

\begin{figure}[tbp!]
\includegraphics[width=0.48\textwidth]{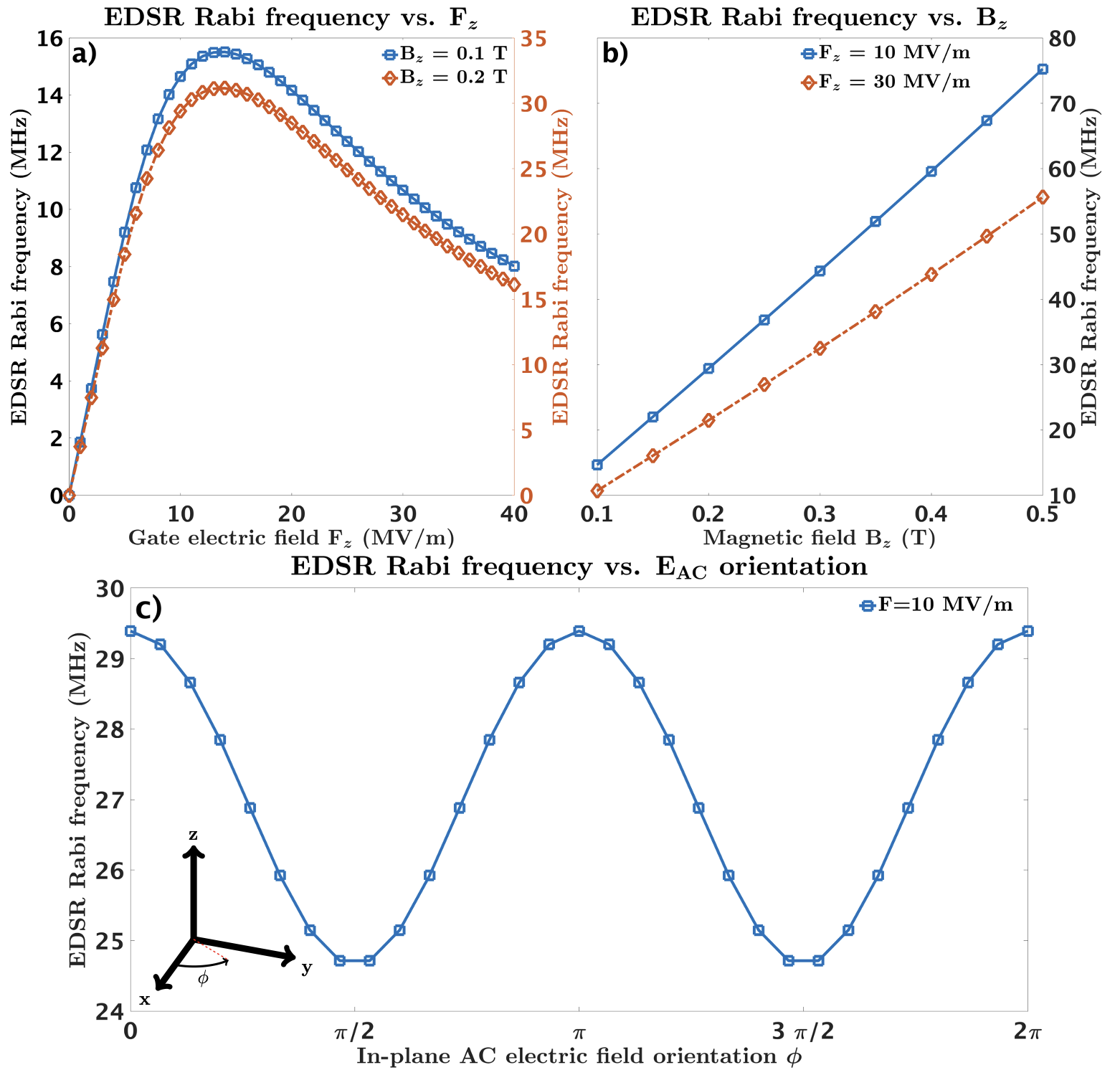}
\caption{\textbf{EDSR Rabi frequency for an out-of-plane magnetic field $\bm{\mathrm{B}_z}$}. 
a) The EDSR Rabi frequency is plotted as a function of the gate electric field $F_z$ for two different out-of-plane magnetic field strengths: B${}_z$\,=\,0.1\,T (solid line with square markers) and B${}_z$\,=\,0.2\,T (dashed line with diamond markers). 
b) The EDSR Rabi frequency is shown as a function of the out-of-plane magnetic field $B_z$ for two different gate electric field strengths: F${}_z$\,=\,20\,MV/m (solid line with square markers) and F${}_z$\,=\,30\,MV/m (dashed line with diamond markers). In this case, an in-plane AC electric field of 1 kV/m is applied.
c) The EDSR Rabi frequency is plotted against the angle of the applied in-plane E${}_{\text{AC}}$ alternating driving electric field. The magnetic field is along the $\hat{z}$-direction with magnitude B${}_z$\,=\,0.1\,T. The top gate field is F${}_z$\,=\,10\,MV/m.
}
\label{fig: EDSR OP All}
\end{figure}

For the purposes of this discussion we have chosen to focus on defects whose electric fields lie primarily in the plane of the qubit. This is because, as will emerge below, regardless of the orientation of the magnetic field, the qubit Larmor frequency exhibits extrema as a function of the top gate electric field. By operating the qubit at these extrema one can protect against fluctuations in the out-of-plane electric field. Thus fluctuations in the in-plane electric field are most detrimental to the qubit, and this is what our model focuses on.

The potential of a single defect can be modelled as \cite{Bermeister2014, Wang2021}:
\begin{equation}\label{Eq - Dephasing potential single charge 1}
U_{\text{scr}}(q)=\frac{e^2}{2 \epsilon_0 \epsilon_r} e^{-q d}\frac{\Theta\left(2 k_F-q\right)}{q+q_{T F}}.
\end{equation}
where $q$ is the wave-vector, $q_{\text{TF}}$ is the Thomas-Fermi wave-vector for silicon, which is independent of the density of holes, and $k_F$ is the Fermi wave vector. The relevant values can be found in Table\,\ref{TB1 - Silicon VS Germanium}. $\Theta$ is the Heaviside step function. In position space, the single charge defect potential can be written as \cite{Bermeister2014, Wang2021}
\begin{equation}\label{Eq - Dephasing potential singler charge 2}
U_{s}(\bm{r})=\frac{e^2}{4 \pi \epsilon_0 \epsilon_r} \frac{1}{q_{\text{TF}}^2}\pqty{ \frac{1}{ \norm{\bm{r}-\bm{r}_D}^3 } } 
\end{equation}
where $\bm{r}_D\,=\,(x_D, y_D, z_D)$ is the position vector of the single charge defect, taking the center of the quantum dot as the origin. For silicon, the relative electrical permeability is $\epsilon_r\,=\,11.68$; $\epsilon_{0}$ is the vacuum electrical permeability. The defect location $x_D$ is set to be 30\,nm from the center of the quantum dot. The resulting change in the dot's orbital splitting (the energy difference between the orbital ground state and first excited state) due to the defect is $1 \mu$eV at F${}_z$\,=\,1 MV/m, consistent with Refs.\,\cite{Connors2022,Connors2019}.
\begin{figure}[tbp!]
\includegraphics[width=0.48\textwidth]{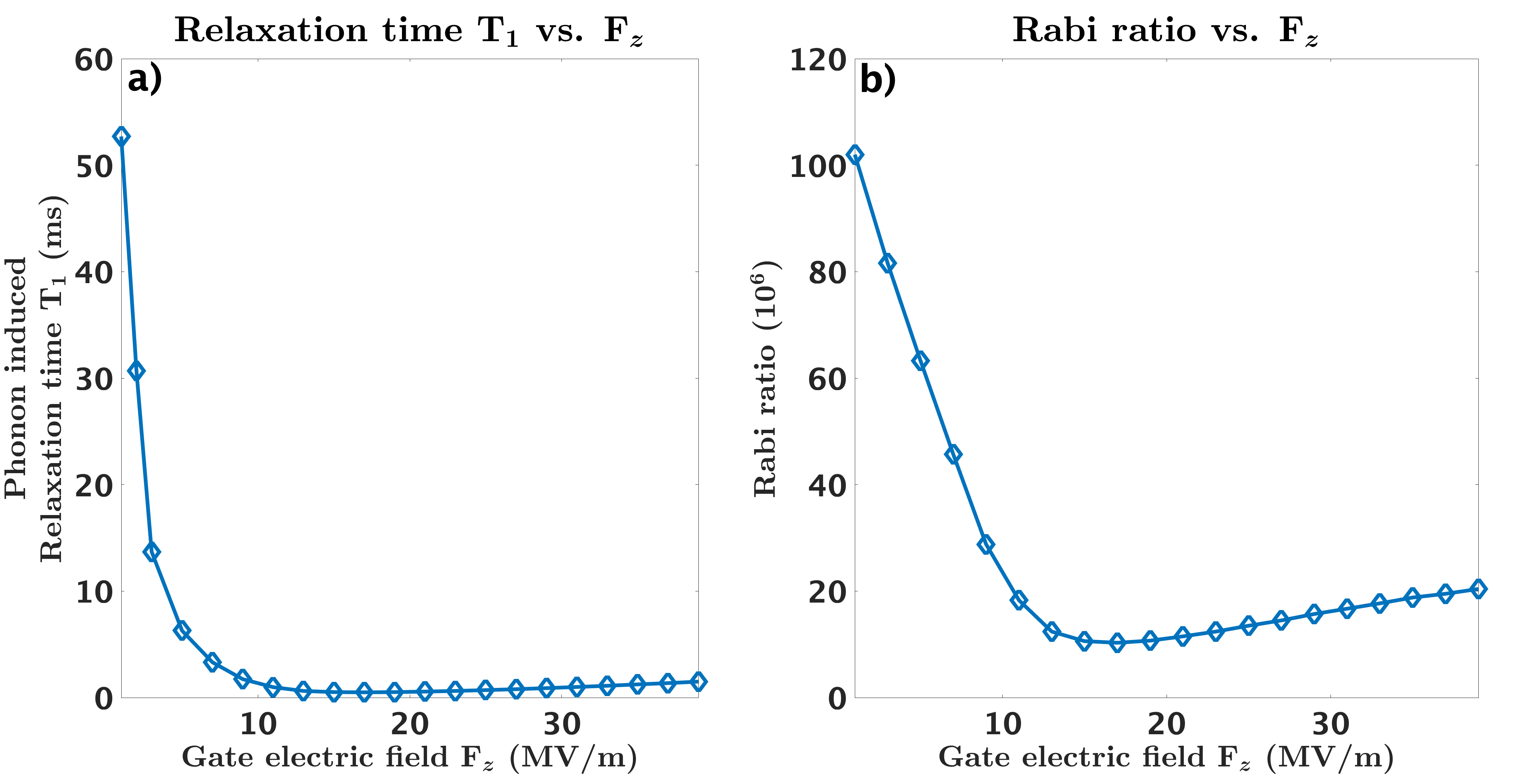}
\caption{\textbf{Relaxation time and Rabi ratio for an out-of-plane magnetic field $\bm{\mathrm{B}_z}$}. 
a) The single phonon relaxation time is calculated for an out-of-plane magnetic field B${}_z$\,=\,0.1\,T. This relaxation time represents the characteristic time scale for the decay of the qubit due to phonon-hole interactions. 
b) The Rabi ratio is plotted as a function of the top gate electric field $F_z$. The Rabi ratio is $\approx 10^6$, which provides an indication of the efficiency of the qubit operation, with higher values indicating a stronger qubit coherence.
}
\label{fig: RR OP All}.
\end{figure}
We also include dipole charge defects, with
\begin{equation}\label{Eq - Dephasing potential Dipole charge}
U_{\mathrm{dip}}\left(\bm{r}\right)=\frac{ \bm{p} \cdot (\bm{r}-\bm{r}_{D}) }{4 \pi \epsilon_0 \epsilon_{\mathrm{r}} (\bm{r}-\bm{r}_D)^3 }
\end{equation}
The dipole moment is $\bm{p}\,=\,e \bm{l}$, where $\bm{l}$ is the dipole vector. In our calculations, we assume the size of the dipole is 0.1\,nm. The influence of the dipole charge defects on the dephasing time is typically much smaller than the influence due to single charge defects \cite{Wang2021,Culcer2013}. By considering both single charge defect and dipole charge defects, we can calculate the dephasing time in the quasi-static limit that estimates the upper bound of the dephasing time, denoted $T_2^*\,=\,2 \pi / \delta\omega$.

We note that this analysis is applicable only to a single defect giving rise to random telegraph noise. The realistic but more complicated case of $1/f$ noise for multiple defects will be considered in a different study.

\section{Results and Discussion}
\label{Result and Discussion}

In this section, we present the main findings of our numerical diagonalization. The section is divided into three parts, which focus on qubit dynamics in an out-of-plane magnetic field, dynamics in an in-plane magnetic field, and $g$-factor anisotropy respectively. 

Our numerical model has identified extreme values of the qubit Zeeman splitting as a function of the top gate field for various parameters in different magnetic field orientations. This is because the vertical electric field creates two opposing Stark effects: it simultaneously increases the HH-LH gap and enhances the HH-LH coupling. The sweet spot is the point where these two sources of Stark shift cancel each other. At this point the variation of the qubit Zeeman splitting vanishes in the first order as a function of the gate electric field, as shown in  Fig.\,\ref{fig: QZS OP All} and Fig.\,\ref{fig: QZS IP All}. This non-linear behavior of the qubit Zeeman splitting leads to similar non-linearities in other important properties of the silicon hole spin qubits.

\begin{figure}[tbp!]
\includegraphics[width=0.48\textwidth]{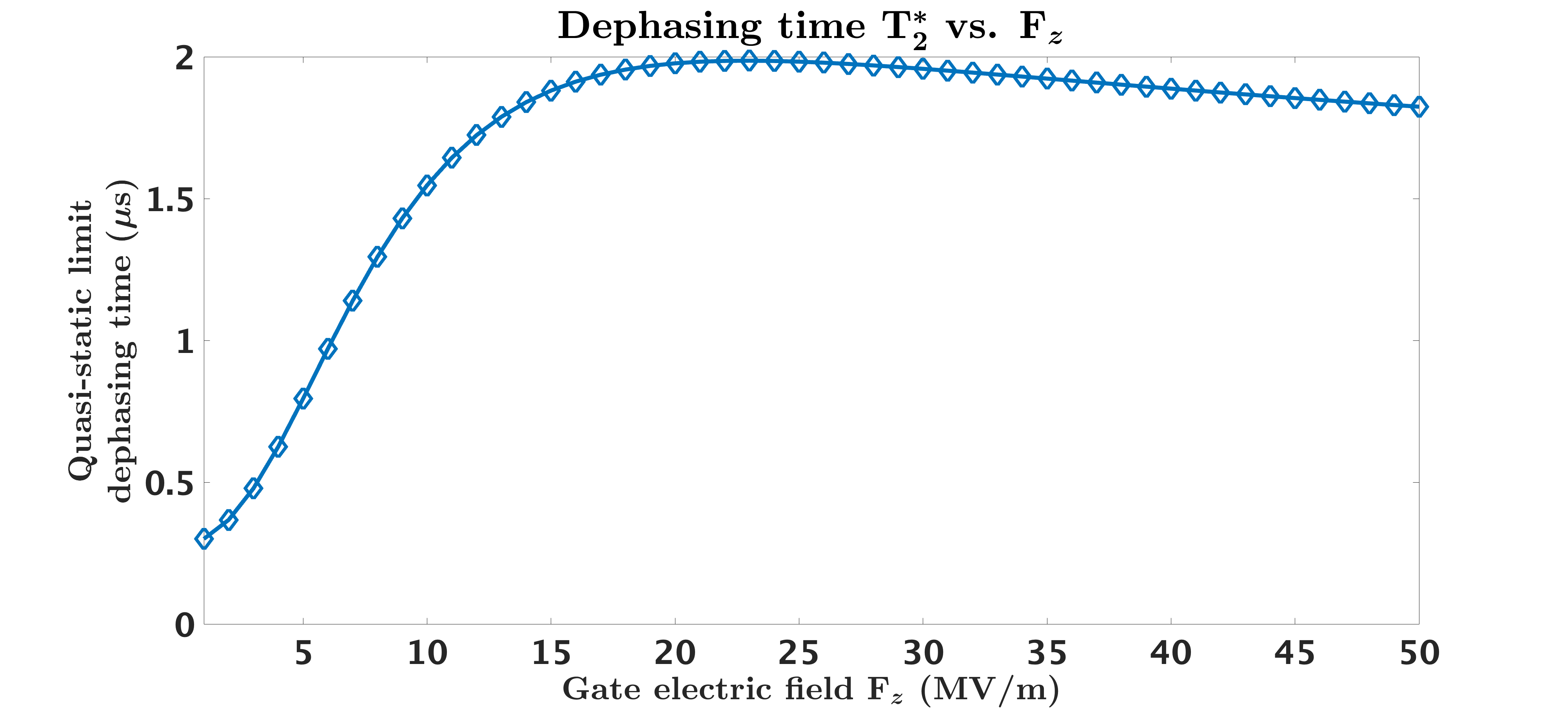}
\caption{\textbf{Dephasing time in the quasi-static limit in an out-of-plane magnetic field $\bm{\mathrm{B}_z}$}. Both single charge defects and dipole charge defects are taken into account, but the dominant contribution to the dephasing potential comes from single charge defects. The dephasing time, which is estimated to be around 2\,$\mu$s, exhibits a local maximum, suggesting the presence of an optimal operation point.}
\label{fig: DP OP F} 
\end{figure}

\subsection{Out-of-plane magnetic field}

The qubit Zeeman splitting $\Delta E_z$ as a function of the gate electric field F$_z$ and the out-of-plane magnetic field B${}_z$ is plotted in Fig.\,\ref{fig: QZS OP All}. When B${}_z$\,=\,0.1\,T, the electric field can change the qubit Zeeman splitting by about $30\%$, and there is a local minimum in the qubit Zeeman splitting as a function of top gate field. 

Next we discuss the EDSR Rabi frequency, shown in Fig.\,\ref{fig: EDSR OP All}. We found that with an in-plane AC electric field of $\sim 1$\,kV/m applied along the $\hat{x}$-direction, the EDSR Rabi frequency is about 50\,MHz at B${}_z$\,=\,0.1\,T. There also exists a local maximum of EDSR frequency as a function of the gate electric field, and the EDSR Rabi frequency is linear in B${}_z$. With an out-of-plane field, the Zeeman splitting term is typically of the order of $\mu$eV, which can be treated perturbatively. Therefore, when B${}_z$ is small, the EDSR Rabi rate can be expanded as a function of B$_z$, with the leading order $\propto$B${}_z$, which is similar to the finding of Ref.\,\cite{Terrazos2021} for Ge. Interestingly, as shown in Fig.\,\ref{fig: EDSR OP All}c, the EDSR Rabi frequency exhibits a slight anisotropy as the electric field is rotated in the plane, varying in magnitude by $\sim 20\%$. This is traced to the presence of the cubic-symmetry $\gamma_3$ terms in the Luttinger Hamiltonian.

\begin{figure}[tbp!]
\includegraphics[width=0.48\textwidth]{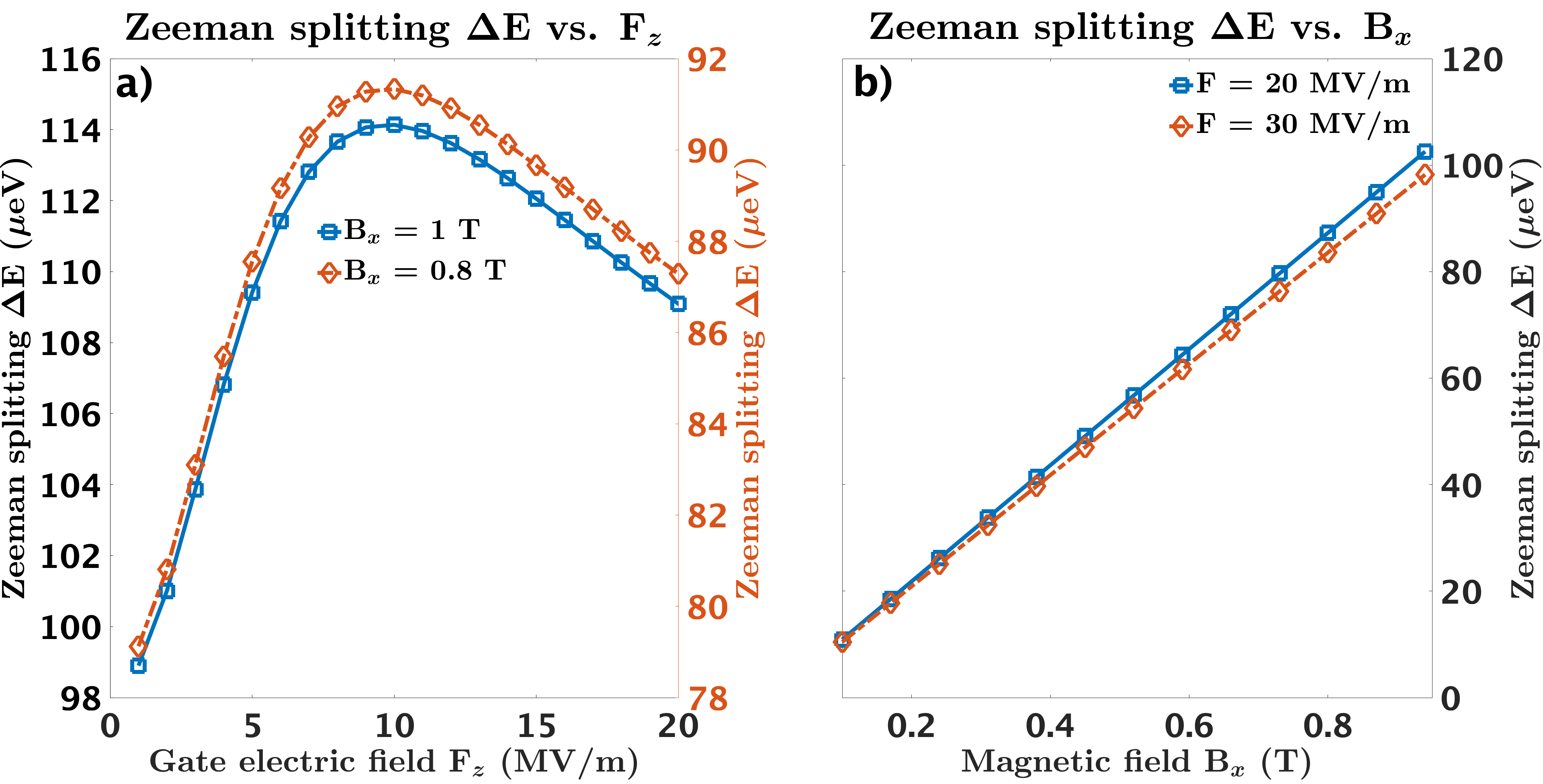}
\caption{\textbf{Qubit Zeeman splitting for an in-plane magnetic field $\bm{\mathrm{B}_\parallel}$}. 
a) The qubit Zeeman splitting is plotted as a function of the gate electric field for two different in-plane magnetic field strengths: B${}_x$\,=\,1\,T (solid line with square markers) and B${}_x$\,=\,0.8\,T (dashed line with diamond markers). Notably, there is a flat local maximum observed around F${}_z$\,=\,11\,MV/m. 
b) The qubit Zeeman splitting is shown as a function of the in-plane magnetic field B${}_x$\,=\,1\,T for two different gate electric field strengths: F${}_z$\,=\,20\,MV/m (solid line with square markers) and F${}_z$\,=\,30\,MV/m (dashed line with diamond markers).
}
\label{fig: QZS IP All} 
\end{figure}

To study the number of operations allowed in one relaxation time, we plot the phonon-induced relaxation time and the Rabi ratio in Fig.\,\ref{fig: RR OP All}. At B${}_z$\,=\,0.1\,T the relaxation time is several milliseconds, which allows $\sim 10^6$ operations. The long relaxation time reflects the weak hole-phonon interactions for silicon as a lighter material with a fast phonon propagation speed compared with germanium. The relaxation rate $\propto B^5$, which is consistent with Ref.\,\cite{Venitucci2020}.

Our results Fig.\,\ref{fig: QZS OP All} and Fig.\,\ref{fig: QZS IP All} show that extrema in the qubit Zeeman splitting as a function of the top gate field F${}_z$ exist for all magnetic field orientations. Nevertheless, proper coherence sweet spots only exist for an out-of-plane magnetic field, as shown in Fig.\,\ref{fig: DP OP F}. To understand this we must highlight the difference between extrema in the qubit Zeeman splitting and $T_2^*$ sweet spots. The qubit states are Kramers conjugates. Time-reversal symmetry implies that the matrix elements giving rise to pure dephasing, which represent an energy difference between the up and down spin states, must involve the magnetic field -- they cannot come from charge noise and spin-orbit coupling alone. Because the magnetic field enters the qubit states both through the Zeeman and the orbital terms, the composition of the qubit states is different depending on the magnetic field orientation. For an out-of-plane magnetic field, the in-plane and out-of-plane dynamics can be approximately decoupled. The main effect of the gate electric field is to give rise to Rashba-like terms acting on the heavy hole spins. Unlike Ge, the Schrieffer-Wolff approximation is not applicable to Si as studied by \cite{Marcellina2017} so this decomposition is not as easily visualized, although one can still envisage an effective spin-orbit Hamiltonian characterized by a Rashba constant. The Rashba term affects the qubit Zeeman splitting, and is directly susceptible to charge noise perpendicular to the interface, which is the main way a defect affects qubit spin dynamics. An in-plane electric field does not couple to the diagonal qubit matrix elements to leading order and can thus be disregarded in a first approximation. 

\begin{figure}[tbp!]
\includegraphics[width=0.48\textwidth]{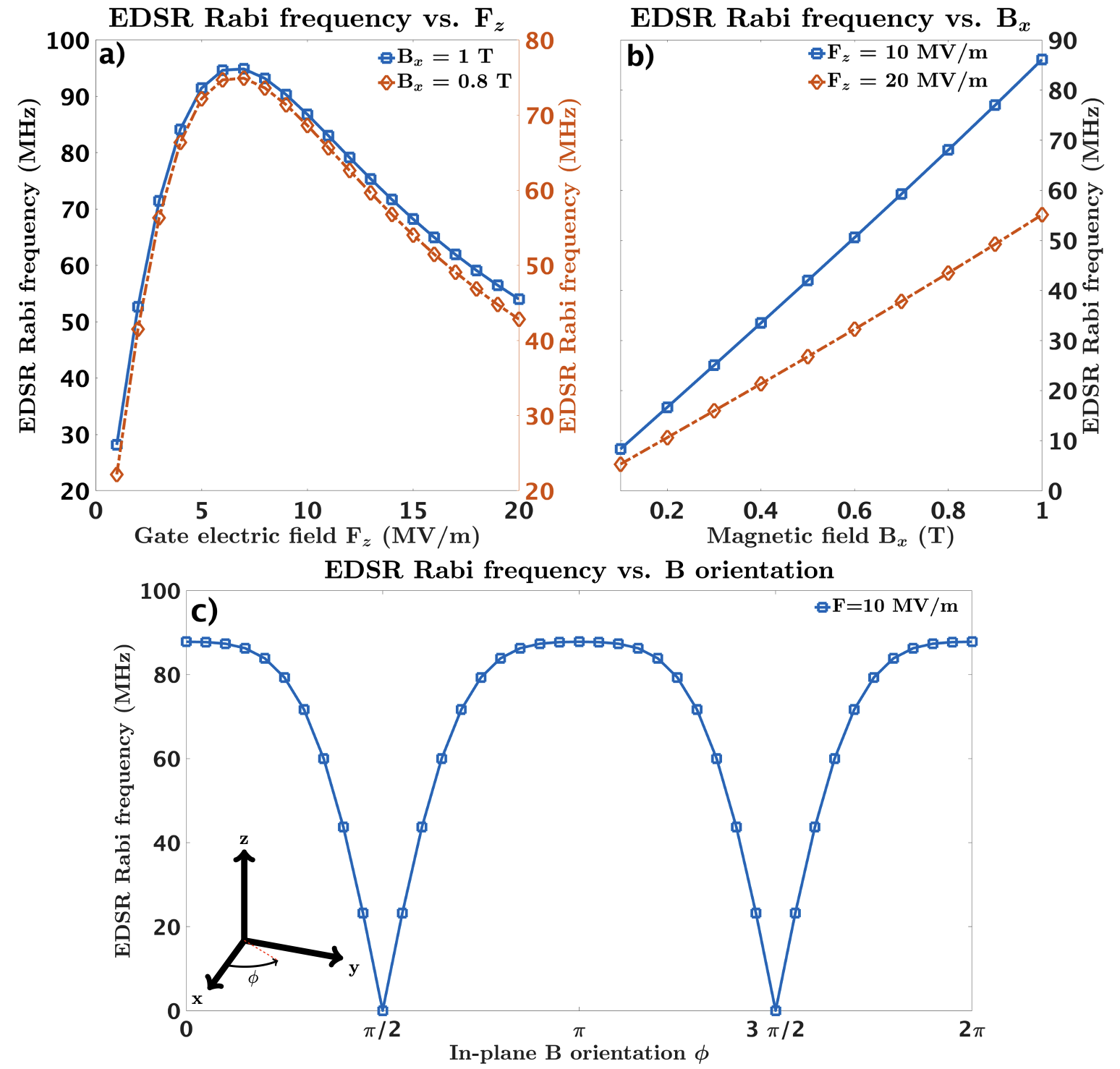}
\caption{\textbf{EDSR Rabi frequency for an in-plane magnetic field $\bm{\mathrm{B}_\parallel}$}. 
a) The EDSR Rabi frequency is plotted as a function of the gate electric field $F_z$ for two different in-plane magnetic field strengths: B${}_x$\,=\,1\,T (solid line with square markers) and B${}_x$\,=\,0.8\,T (dashed line with diamond markers).
b) The EDSR Rabi frequency is shown as a function of the in-plane magnetic field B${}_x$\,=\,1\,T for two different gate electric field strengths: F${}_z$\,=\,10\,MV/m (solid line with square markers) and F${}_z$\,=\,20\,MV/m (dashed line with diamond markers).
c) The EDSR Rabi frequency is plotted as a function of the in-plane magnetic field orientation. In this case, the magnitude of the in-plane magnetic field is fixed at 1 T, while the in-plane AC electric field remains at 1\,kV/m. The magnetic field is rotated through $2\pi$ in the $xy$-plane. Notably, the maximum EDSR Rabi frequency occurs when the magnetic field is aligned along the $\hat{x}$-direction, which coincides with the direction of the in-plane AC driving electric field.
}
\label{fig: EDSR IP All}
\end{figure}

\subsection{In-plane magnetic field}

When magnetic field is applied in the plane, our result (Fig.\,\ref{fig: QZS IP All}) shows a large qubit Zeeman splitting variations as a function of the top gate field, which is also observed in a recent experiments (Ref.\,\cite{Liles2021}). A local minimum as a function of the gate electric field continues to exist, however, in an in-plane magnetic field the local minimum in the qubit Zeeman splitting does not protect the qubit from the single charge defect noise, as was emphasized above in the out-of-plane magnetic field case.

The EDSR Rabi frequency in an in-plane field contains a dominant term $\propto B_x$ as well as a small distortion $ \propto B_x^2$ due to the orbital magnetic field terms. For an in-plane field, the orbital term in Luttinger-Kohn Hamiltonian can change the dispersion of holes significantly, away from the center of the Brillouin zone, the orbital term will distort the parabolicity of the dispersion. Therefore, an in-plane magnetic field will eventually result in stronger heavy-hole-light-hole mixing and significant modulation of the $g$-factor even when the amplitude of the field is small. Considering the transition matrix element in Eq.\,\ref{Eq - EDSR}, this amplitude is determined by the in-plane AC electric field as well as the shapes of the ground state wave function $\ket{\mathbb{0}}$ and of the excited state wave function $\ket{\mathbb{1}}$. As a result, the EDSR Rabi frequency exhibits a non-linear behavior as a function of $B_x$ due to the heavy-hole-light-hole admixture and due to the orbital magnetic field terms. We also observe a strong anisotropy in the EDSR Rabi frequency: applying the magnetic field parallel to the AC in-plane electric field results in an enhanced EDSR Rabi frequency, as shown in Fig.\,\ref{fig: EDSR IP All} c).

\begin{figure}[tbp!]
\includegraphics[width=0.48\textwidth]{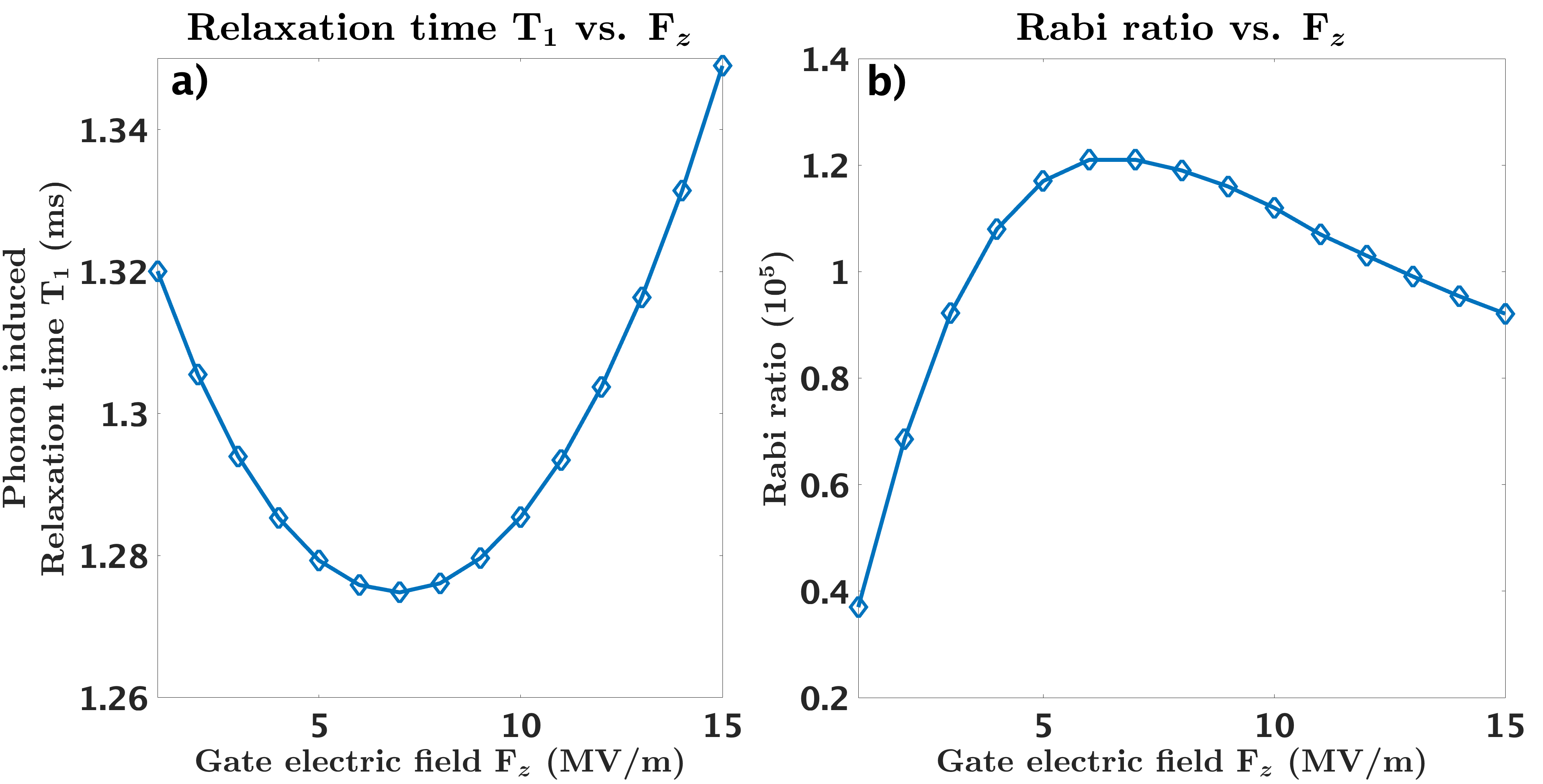}
\caption{\textbf{Relaxation time and Rabi ratio for an in-plane magnetic field $\bm{\mathrm{B}_\parallel}$}. 
a) The single phonon relaxation time is calculated for an in-plane magnetic field B${}_x$\,=\,1\,T. 
b) The Rabi ratio is plotted as a function of the gate electric field F${}_z$. The Rabi ratio, which is approximately around $10^5$.
}
\label{fig: RR IP All} 
\end{figure}

\begin{figure}[tbp!]
\includegraphics[width=0.48\textwidth]{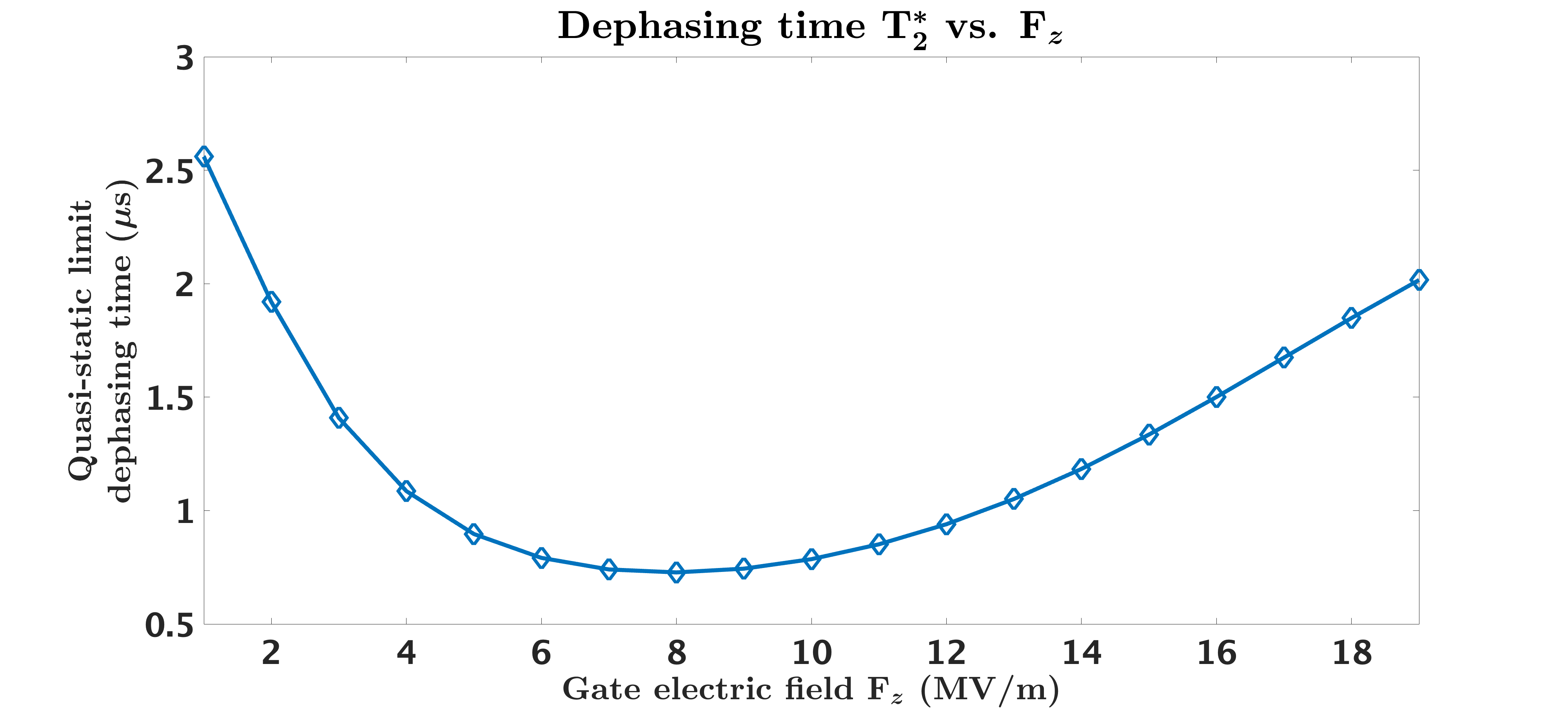}
\caption{\textbf{Dephasing time in quasi-static limit for an in-plane magnetic field $\bm{\mathrm{B}_\parallel}$}. Both single charge defects and dipole charge defects are taken into account, but the dominant contribution to the dephasing potential still comes from single charge defects. The dephasing time exhibits a local minimum, suggesting the absence of an optimal operation point.
}
\label{fig: DP IP F}
\end{figure}

For an in-plane magnetic field, the orbital vector potential terms couple the in-plane and out-of-plane dynamics and no separation of the dynamics is possible. The net effect of this is that the qubit is sensitive to all components of the defect electric field, and an extremum in the qubit Zeeman splitting as a function of the perpendicular electric field does not translate into a coherence sweet spot for charge noise, as Fig.\,\ref{fig: DP IP F} shows.

\subsection {Ellipticity and in-plane \texorpdfstring{$\bm{g}$}{g}-factor anisotropy}

\begin{figure}[tbp!]
\includegraphics[width=0.48\textwidth]{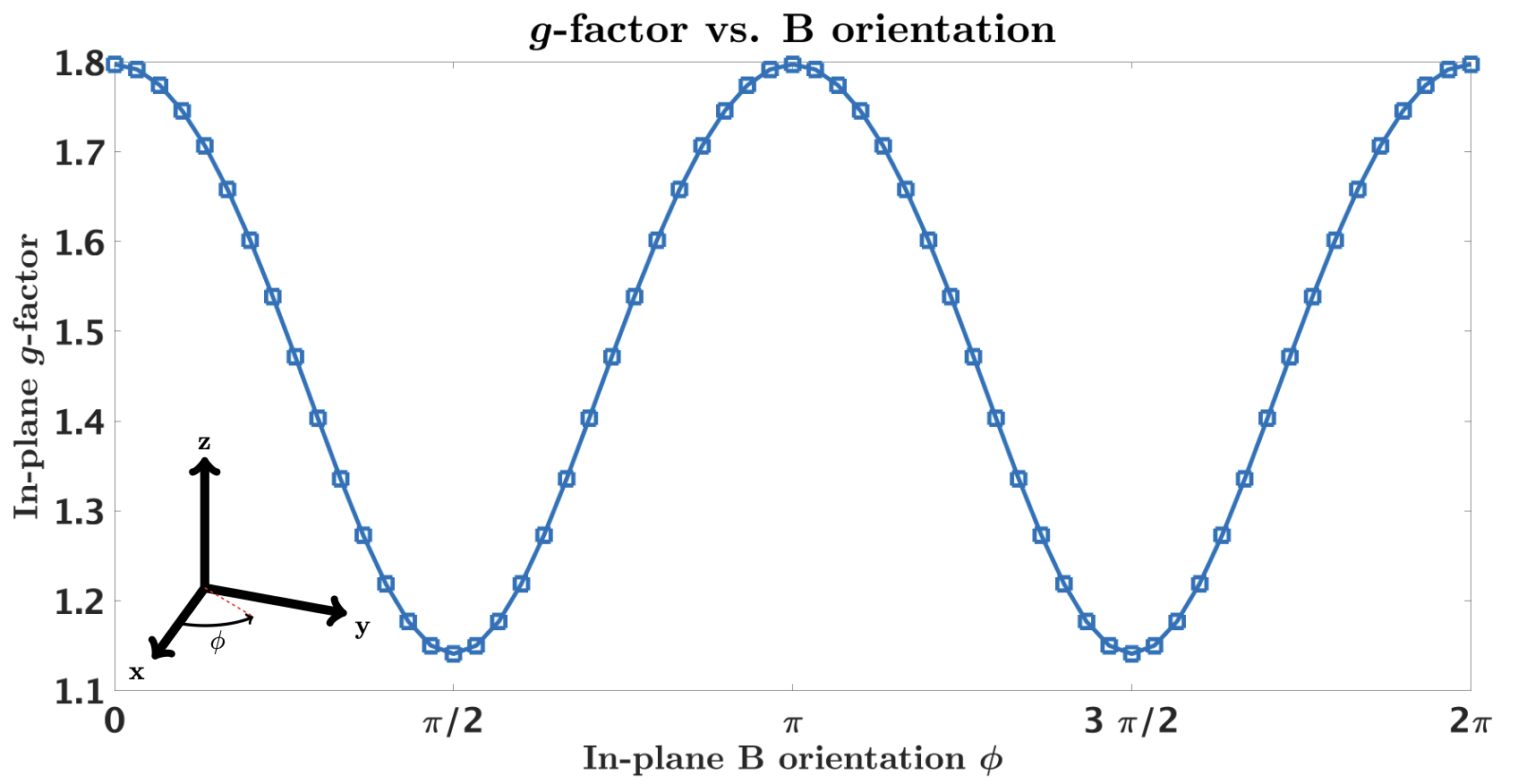}
\caption{\textbf{$g$-factor anisotropy}. The variation of the $g$-factor is plotted as a function of the in-plane magnetic field orientation $\phi$. Here, $\phi$ represents the angle of the magnetic field with respect to the $\hat{x}$-direction. The $g$-factor is determined for an in-plane magnetic field with magnitude of 1\,T. The semi-major axis of the dot is $a_{x}$\,=\,24\,nm; the semi-minor axis of the dot is $a_{y}$\,=\,20\,nm, giving the aspect ratio to be 1.2. When the magnetic field is parallel to the semi-major axis ($\phi$\,=\,$0^{\circ}$), the $g$-factor has a maximum value, which is also observed in Ref.\,\cite{Liles2021}.
}
\label{fig: g-factor Anisotropy}
\end{figure}

In experimental studies, dots are often formed without explicitly attempting to remain circular, leading to a notable anisotropy in the effective $g$-factors, as depicted in Fig.\,\ref{fig: g-factor Anisotropy}. Despite the presence of an anisotropy term, denoted as $\kappa_2$ in the Zeeman Hamiltonian, it is important to note that $\kappa_2$ is typically smaller than $\kappa_1$ in both group IV and group III-V hole systems. Therefore, it has limited impact on the energy spectrum of the qubit. In contrast, the orbital term in $H_{\text{{LK}}}$ and the effective mass will contribute more strongly to the anisotropy of the Rashba spin-orbit coupling and of the effective $g$-factors. Additionally, while the linear Rashba term assumes a central role in circular dots, the cubic Rashba term becomes activated in elliptical quantum dots, resulting in enhanced Rashba spin-orbit coupling and faster EDSR Rabi oscillations. 

For possible experimental settings, the lateral confinement in the $\hat{x}$ and $\hat{y}$ directions can be independently adjusted using the electrostatic gates. This corresponds to in-situ control over $\omega_{x,y}$, which are defined in Eqs.\,\ref{Eq - x wave functions 3}-\ref{Eq - y wave functions 4}. Previous work by Qvist and Danon (Ref.\,\cite{Qvist2022}) investigated lateral confinement potentials, providing an analytical study of effective mass anisotropy and the size of the confinement potential by taking the linear Rashba term as an example in a perturbative approach on the four-band Luttinger-Kohn Hamiltonian. In contrast, our numerical calculations include all Rashba terms, involving tracing all non-commutable canonical momentum operators in higher excited states, and accounting for the non-parabolic behaviors of the band structure based on a six-band Luttinger-Kohn Hamiltonian. 

Our results indicate that the $g$-factor exhibits an oscillating pattern when we rotate a constant in-plane magnetic field in the $xy$-plane. For example, when the aspect ratio is 1.2, we observe a $g$-factor variation up to 100\% as a function of the in-plane magnetic field angle. This substantial anisotropy in the in-plane $g$-factor is consistent with recent experimental observations (Ref.\,\cite{Liles2021}). 

\section{Comparison between Germanium and Silicon}
\label{Comparison between Germanium and Silicon}
A promising competitor for silicon, in semiconductor quantum dot hole spin qubit area, is germanium. However, due to the fabrication details of silicon hole quantum dot and germanium hole quantum dot, the location of the sweet spot as a function of the top gate field, the strain in the sample, and the modulation of the in-plane and out-of-plane $g$-factors are expected to be different. As a comparison of the material characters, we list important parameters, which is relevant in fabricating the hole spin qubit, of silicon and germanium in Table.\,\ref{TB1 - Silicon VS Germanium}.

The in-plane effective mass of a hole in silicon (0.216$m_0$) is much heavier than that in germanium (0.057$m_0$). As a consequence, the heavy-hole-light-hole energy splitting in silicon (around 5\,meV) will be much smaller than that in germanium (around 50\,meV). A direct outcome of a small heavy-hole-light-hole energy splitting is that, the presence of strains will efficiently lead to mixing between the light-hole and heavy-hole band, which will amplifies the Stark shift effect. Experimental data indicates that the in-plane $g$-factor in silicon can range between 1.5-2.5 \cite{Liles2021}, whereas in germanium hole quantum dots, the $g$-factor only exhibits small variations 0.16-0.3 \cite{Hendrickx20202, Hendrickx2021}. Furthermore, the smaller effective mass in silicon imposes limitations on the splitting of quantum dot orbital levels, thereby restricting the size of silicon hole quantum dots.

\begin{figure}[tbp!]
\includegraphics[width=0.48\textwidth]{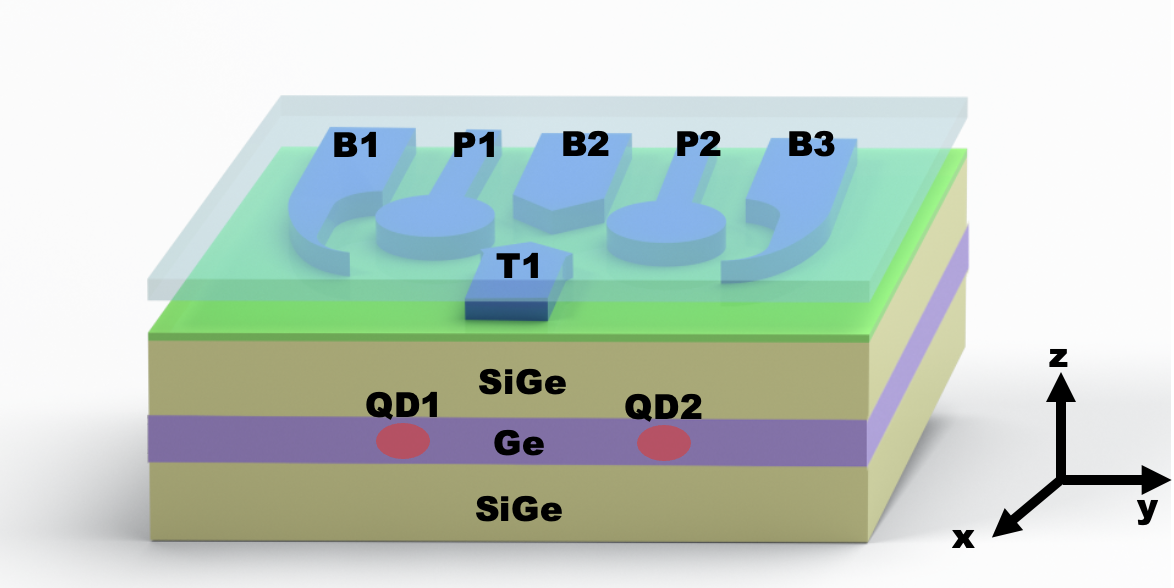}
\caption{\textbf{A prototype double quantum dot in strained germanium hole system}. The strained germanium quantum well is grown epitaxially on a strain-relaxed SiGe layer. The portion of the silicon in SiGe can be tuned experimentally, therefore, the axial strain can be controlled. Gate B2 and T1 can control the inter-dot tunneling. In this picture, we set the growth direction to be along $\hat{z}$-direction.
}
\label{fig: Germanium Qubit Prototype}
\end{figure}

The strain present in both silicon and germanium hole quantum dots is another determinant factor explaining the different $g$-factor modulations and the Rabi ratios. In general, axial strain terms (i.e., $P_\varepsilon$, $Q_\varepsilon$, $P_\varepsilon+Q_{\varepsilon}$, $P_\varepsilon-Q_{\varepsilon}$ in $H_{\text{LKBP}}$) will change the the heavy-hole-light-hole energy splitting directly, while shear strain terms (i.e., $R_\varepsilon$, $S_\varepsilon$ in $H_{\text{LKBP}}$) intermix the heavy-hole and light-hole states. 

For silicon hole quantum dots based on metal-oxide-semiconductor platforms, strain naturally develops in the device due to the differences in the thermal contraction between metal electrodes and the silicon substrate. While strain engineering is a common practice in the classical electronics industry, academic research into quantum dots has not been focused on this aspect thus far, except for the occasional consideration on choices of material stacks \cite{Thorbeck2015}.
In the case of germanium hole spin qubits based on a homogeneous uni-axial strain strained germanium quantum well in a SiGe/Ge/SiGe heterostructure, strain can be meticulously controlled. The substrate includes a fully strain-relaxed SiGe layer. The middle of the heterostructure comprises an epitaxially grown layer of strained germanium, hosting the hole qubit, and another layer of relaxed SiGe atop the Ge layer. The concentration of Si atoms in Si$_x$Ge$_{1-x}$, represented by the component fraction factor $x$, also determines the strain in the pure Ge layer via Vegard’s law. To quantitatively compare the strain in the silicon and the germanium hole spin qubit devices, we use typical parameters as listed in Ref.\,\cite{Terrazos2021}. For instance, if the relaxed SiGe layer is Si$_{0.25}$Ge$_{0.75}$, ($x$\,=\,0.25) the axial compressive strain will be $\epsilon_{xx}$\,=-\,0.001, which is $\sim 5$ times larger than the strain present in the silicon metal-oxide-semiconductor quantum dot. In Table.\,\ref{TB2 - Silicon VS Germanium2}, we summarize various typical configurations, including strains, top gate fields, magnetic fields, and geometries to reach the optimal operation points in different materials. We notice that the parameters used to fit experimental data, such as the dot geometry, shear strain, and axial strain, are estimates. It is crucial to include the non-uniform strain from the gate electrodes, as shown in Ref.\cite{Liles2021}. For more precise results, direct strain profilling as in Ref.\,\cite{stehouwer2023}, or device-specific modelling can be employed. Strain will be thoroughly investigated in future works. In this context, we note that we do not anticipate strain to change the existence of the optimal operation points of the qubits for fast EDSR Rabi ratio and minimized dephasing time as a function of the top gate field.

Another important difference between Ge and Si concerns the applicability of the Schrieffer-Wolff transformation in analyzing qubit Hamiltonians. For Ge, a perturbative approach based on the Schrieffer-Wolff transformation is demonstrated to be effective for an out-of-plane magnetic field,\cite{Wang2021} which relies on the large heavy-hole-light-hole splitting in a low density Ge system. In Si the heavy-hole-light-hole splitting is much smaller than in Ge, while the cubic-symmetry term in the Luttinger Hamiltonian is very strong. As a result of this, the Schrieffer-Wolff transformation cannot account for full density-dependence (i.e., quantum dot radius dependence) of the hole states and split-off band correctly, and a full diagonalization of the Hamiltonian is needed to yield accurate results.

\begin{table}[tbp!]
\caption{\textbf{Comparison of silicon and germanium material parameters}. The relevant parameters defining the silicon hole spin qubit and germanium hole spin qubits are collected from Ref.\,\cite{Wortman1965,Hopcroft2010,Hao2000,Soline2004,osten1987}. In the table, the out-of-plane heavy-hole band and light-hole band mass is defined as $m_{\text{HH}}=m_0/\pqty{\gamma_1-2\gamma_2}$, $m_{\text{LH}}=m_0/\pqty{\gamma_1+2\gamma_2}$; the in-plane heavy-hole band effective mass and the light-hole band effective mass is defined as $m_{\text{HP}}\,=\,m_0/\pqty{\gamma_1+\gamma_2}$, $m_{\text{LP}}\,=\,m_0/\pqty{\gamma_1-\gamma_2}$. $m_0$ is the bare electron mass, $\gamma_1$, $\gamma_2$, $\gamma_3$ are Luttinger parameters. The density $\rho$ is the bulk density of isotropic silicon or germanium. $v_{\ell}$, $v_{t_1}$, $v_{t_2}$ are phonon propagation speeds along three different polarizations. $a_v$, $b_v$, $d_v$ are the hydro-static deformation potential constant, uni-axial deformation potential constant, and shear deformation potential constant respectively. The split-off band gap is denoted by $\Delta_0$.
}
\label{TB1 - Silicon VS Germanium}
\begin{ruledtabular}
\begin{tabular}{ccc}
\textrm{Parameters}&
\textrm{Silicon}&
\multicolumn{1}{c}{\textrm{Germanium}}\\
\colrule
$\gamma_1$ & 4.29 & 13.38 \\
$\gamma_2$ & 0.34 & 4.24 \\
$\gamma_3$ & 1.45 & 5.69 \\
$\kappa_1$ & -0.42 & 3.41 \\
$\kappa_2$ & 0.01 & 0.06 \\
$m_{\text{HH}}$ & 0.277 $m_0$ & 0.204 $m_0$ \\
$m_{\text{LH}}$ & 0.201 $m_0$ & 0.046 $m_0$ \\
$m_{\text{HP}}$ & 0.216 $m_0$ & 0.057 $m_0$ \\
$m_{\text{HP}}$ & 0.253 $m_0$ & 0.109 $m_0$ \\
$\rho$ & 2329 kg/$\text{m}^3$ & 5330 kg/$\text{m}^3$ \\
$v_\ell$ & 9000 m/s & 3570 m/s \\
$v_{t_1}$ & 5400 m/s & 4850 m/s \\
$v_{t_2}$ & 5400 m/s & 4850 m/s \\
$a_v$ & 2.38 eV & 2.00 eV \\
$b_v$ & -2.10 eV & -2.16 eV\\
$d_v$ & -4.85 eV & -6.06 eV\\
$\Delta_0$ & 44 meV & 296 meV \\
\end{tabular}
\end{ruledtabular}
\end{table}

\begin{table}[tbp!]
\caption{\textbf{Possible configurations for optimal operation points}. In this table, the diameter of the quantum dot is $d_x\,=\,d_y\,=\,$40\,nm, and the width of the quantum well is L\,=\,13\,nm. We consider both in-plane and out-of-plane magnetic fields for silicon and germanium. The strains $\varepsilon_{xx}$, $\varepsilon_{yy}$ and $\varepsilon_{zz}$ have the relation $\varepsilon_{xx}\,=\,\varepsilon_{yy}$\,=\,$\varepsilon_{zz}\,=\,-\varepsilon_{xx} (2 C_{12}/ C_{11})$, the shear strain $\varepsilon_{xz}$ is estimated from Ref.\,\cite{Liles2021}. We use $\Delta E_{\text{Z}}$ to denote the qubit Zeeman splitting. Note that there exist many possible combinations of parameters to get an optimal operation point as a function of the gate electric field.
}
\label{TB2 - Silicon VS Germanium2}
\begin{ruledtabular}
\begin{tabular}{lcc}
\textrm{Confinements}&
\textrm{Silicon}&
\multicolumn{1}{c}{\textrm{Germanium}}\\
\colrule
Orbital energy splitting & 0.3\,meV & 0.3\,meV\\
HH-LH energy splitting & 7\,meV & 100\,meV \\
Typical $\varepsilon_{xx}$ & 0.001 & 0.01 \\
Typical $\varepsilon_{yy}$ & 0.001 & 0.01 \\
Typical $\varepsilon_{zz}$ & -0.00077 & -0.0077 \\
Typical $\varepsilon_{xz}$ & 0.0008 & 0 \\
$\Delta E_{\text{Z}}$ ($B_x\,=\,1$T) & 100\,$\mu$eV & 15\,$\mu$eV \\
Sweet spot ($B_x\,=\,1$T) & 8\,MV/m & 18\,MV/m \\
$\Delta E_{\text{Z}}$ ($B_z\,=\,0.1$T) & 10\,$\mu$eV & 90\,$\mu$eV \\
Sweet spot ($B_z\,=\,0.1$T) & 13\,MV/m & 20\,MV/m \\
\end{tabular}
\end{ruledtabular}
\end{table}

\section{Conclusions and Outlook}
\label{Conclusions}
In this paper, starting from the diagonalization of the Luttinger-Kohn-Bir-Pikus Hamiltonian, we have developed a numerical method to study the silicon hole spin qubits in different experimental configurations. We have shown that the gate electric field significantly modulates the qubit Zeeman splitting, EDSR Rabi frequency and relaxation time. We have shown that the dephasing time due to random telegraph noise stemming from single and dipole charge defects exhibits very different behaviors in in-plane and out-of-plane magnetic fields. We find that in an out-of-plane magnetic field coherence sweet spots can be identified as a function of the top gate field, at which random telegraph noise does not couple to the spin. However, in the case of in-plane fields the role of random telegraph noise can be reduced but not entirely removed, because the vector potential terms expose the qubit to all components of a defect's electric field. The numerical method we have developed in this work can be extended to many-hole spin qubits in other materials as well as to studies of several qubits required for entanglement. 

\begin{table}[tbp!]
\caption{\textbf{Comparison of the EDSR Rabi time, relaxation time and Rabi ratio between silicon and germanium with same orbital energy and qubit Zeeman splittings in an in-plane magnetic field}. The strain used in germanium is $\varepsilon_{yy}$\,=\,$\varepsilon_{xx}$\,=\,0.001, $\varepsilon_{zz}$\,=$-(2C_{12}/C_{11})\varepsilon_{xx}$. The strain used in silicon is $\varepsilon_{xx}$\,=\,$\varepsilon_{yy}$\,=\,0.1\%, $\varepsilon_{zz}$\,=\,-0.077\%, $\varepsilon_{xz}$\,=\,0.08\%, $\varepsilon_{xy}$\,=\,$\varepsilon_{yz}$=0. One can always adjust the gate electric field and the magnetic field to ensure the qubit Zeeman splittings are the same in a silicon and a germanium hole qubit. For EDSR Rabi frequency, the in-plane AC electric field used is 1\,kV/m. To have the same orbital confinement energy to be 0.3\,meV, the silicon dot size is $a_y$\,=\,$a_x$=25\,nm, L\,=13\,nm, while the germanium dot size is $a_y$\,=\,$a_x$\,=50\,nm, L\,=11\,nm. The gate electric field is fixed to be 10\,MV/m.
}
\label{TB3 - Silicon VS Germanium 3}
\begin{ruledtabular}
\begin{tabular}{lcc}
\textrm{Confinements}&
\textrm{Silicon}&
\multicolumn{1}{c}{\textrm{Germanium}}\\
\colrule
Orbital energy splitting & 0.3\,meV & 0.3\,meV\\
$\Delta E_{\text{Z}}=25\,\mu$eV & $B_x=0.3$\,T & $B_x=2$\,T \\
EDSR Rabi time & 80\,ns & 200\,ns \\
Relaxation time & 200\,ms & 4\,ms\\
Rabi Ratio & 2$\times10^6$ & 2$\times10^4$\\
Dephasing time & 0.7\,$\mu$s & 10\,$\mu$s
\end{tabular}
\end{ruledtabular}
\end{table}

\section{Acknowledgments}. 
We are grateful to Tetsuo Kodera for a series of stimulating discussions. This project is supported by the Australian Research Council Centre of Excellence in Future Low-Energy Electronics Technologies (project number CE170100039).

\newpage
\bibliography{Reference}

\clearpage
\renewcommand{\theequation}{S\arabic{equation}}
\renewcommand{\thefigure}{S\arabic{figure}}
\renewcommand{\bibnumfmt}[1]{[S#1]}
\renewcommand{\citenumfont}[1]{S#1}

\appendix

\onecolumngrid
\section{Silicon hole qubit Hamiltonian}
If we consider the Luttinger Hamiltonian along the crystallographic axis [100], [010], [001], the usual $4 \times 4$ valence band structure can be described by the Luttinger Hamiltonian:
\begin{footnotesize}
\begin{align}
H_{\text{LK}}=\frac{\hbar^2}{2m_0}\left[\left(\gamma_1+\frac{5}{2} \gamma_2\right) k^2 \bm{I} - 2 \gamma_2\left(k_x^2 J_x^2+k_y^2 J_y^2+k_z^2 J_z^2\right) - 4 \gamma_3\left(\left\{k_x, k_y\right\}\left\{J_x, J_y\right\}+\left\{k_y, k_z\right\}\left\{J_y, J_z\right\}+\left\{k_z, k_x\right\}\left\{J_z, J_x\right\}\right)\right]
\end{align}
\end{footnotesize}
where $\bm{I}$ is a $4 \times 4$ identity matrix. The curly bracket $\{.,.\}$ denotes the anti-commutator of the quantities defined by $\{A,B\}=(AB+BA)/2$.

We should notice that due to the existence of the vector potential when we quantized the Hamiltonian in all three dimensions, the anti-commutator can not be simply added with each other, the extended six-band Luttinger-Kohn Hamiltonian will read
\begin{equation}
    H_{\text{LK}}\,=\,\mqty[ P+Q & 0 & -S & R & \dsp{-\frac{1}{\sqrt{2}}} S & \sqrt{2} R \\ 0 & P+Q & R^* & S^* & -\sqrt{2} R^* & -\dsp{\frac{1}{\sqrt{2}}} S^* \\ -S^* & R & P-Q & 0 & -\sqrt{2}Q & \dsp{\sqrt{\frac{3}{2}}} S \\ R^* & S & 0 & P-Q & \dsp{\sqrt{\frac{3}{2}}} S^* & \sqrt{2}Q \\ -\dsp{\frac{1}{\sqrt{2}}} S^* & -\sqrt{2}R & -\sqrt{2}Q^* & \dsp{\sqrt{\frac{3}{2}}} S & P+\Delta & 0 \\ \sqrt{2}R^* & -\dsp{\frac{1}{\sqrt{2}}}S & \dsp{\sqrt{\frac{3}{2}}} S^* & \sqrt{2}Q^* & 0 & P+\Delta]
\end{equation}
where the Luttinger-Kohn matrix elements are
\begin{equation}
\begin{aligned}
    P &= \frac{\hbar^{2}}{2 m_{0}} \gamma_{1}\left(k_{x}^{2}+k_{y}^{2}+k_{z}^{2}\right) \\
    Q &= -\frac{\hbar^{2}}{2 m_{0}} \gamma_{2}\left(2 k_{z}^{2}-k_{x}^{2}-k_{y}^{2}\right) \\
    R &= \sqrt{3} \frac{\hbar^{2}}{2 m_{0}}\left[-\gamma_{2}\left(k_{x}^{2}-k_{y}^{2}\right)+2 i \gamma_{3} k_{x} k_{y}\right] \\
    S &= 2\sqrt{3} \frac{\hbar^{2}}{2m_{0}} \gamma_{3}\left(k_{x}-i k_{y}\right) k_{z}
\end{aligned}
\end{equation}
More explicitly:
\begin{equation}
	P+Q\,=\,\frac{\hbar^2}{2m_0}\pqty{\gamma_1-2\gamma_2} k_z^2 + \frac{\hbar^2}{2m_0}\pqty{\gamma_1+\gamma_2}\pqty{k_x^2+k_y^2}
\end{equation}
\begin{equation}
	P-Q\,=\,\frac{\hbar^2}{2m_0}(\gamma_1+2\gamma_2)k_z^2+\frac{\hbar^2}{2m_0}(\gamma_1-\gamma_2)(k_x^2+k_y^2)
\end{equation}
From these equations, we can write down the effective mass directly
\begin{equation}
\begin{aligned}
m_{\mathrm{HH}} & \equiv \frac{m_{0}}{\gamma_{1}-2 \gamma_{2}} \quad m_{\mathrm{LH}} \equiv \frac{m_{0}}{\gamma_{1}+2 \gamma_{2}} \\
m_{\mathrm{HP}} & \equiv \frac{m_{0}}{\gamma_{1} + \gamma_{2}} \quad m_{\mathrm{LP}} \equiv \frac{m_{0}}{\gamma_{1} - \gamma_{2}}
\end{aligned}
\end{equation}

\section{Numerical diagonalization}

In this section, we will discuss the details of the numerical diagonalization of the qubit Hamiltonian. In general, we can just project our qubit Hamiltonian onto the wave functions we are using up to as many levels as we want. However, performing integrals analytically over all the wave functions is time-consuming. However, due to the parity of the wave functions and the symmetry of our confinement potential, we can use the following relations to further simplify the integrals.

If $\phi$ is the simple Harmonic oscillator wave functions, we will have:
\begin{equation}
\int_{-\infty}^{\infty} \phi_{n}(x) \frac{d \phi_{m}}{d x} d x= \begin{cases} a \sqrt{\dsp\frac{n+1}{2}} & m=n+1 \\[0.7em] -a \sqrt{\dsp\frac{n}{2}} & m=n-1 \\[0.7em] 0 & \text { otherwise }\end{cases}
\end{equation}
\begin{equation}
\int_{-\infty}^{\infty} \phi_{m}(x) x \phi_{n}(x) d x= \begin{cases}\dsp\frac{1}{a} \sqrt{\frac{n+1}{2}} & m=n+1 \\[0.7em] \dsp\frac{1}{a} \sqrt{\frac{n}{2}} & m=n-1 \\[0.7em] 0 & \text { otherwise }\end{cases}
\end{equation}
\begin{equation}
\int_{-\infty}^{\infty} \phi_{m}(x) x^{2} \phi_{n}(x) d x= \begin{cases}\dsp\frac{\dsp\sqrt{n(n-1)}}{2a^{2}} & m=n-2 \\[0.7em] \dsp\frac{2 n+1}{2 a^{2}} & m=n \\[0.7em] \dsp\frac{\sqrt{(n+1)(n+2)}}{2 a^{2}} & m=n+2 \\[0.7em] 0 & m \neq n \neq n \pm 2\end{cases}
\end{equation}
When we evaluating the integrals, we need to replace $a$ by appropriate dot radius ($a_x$ or $a_y$). The out-of-plane wave functions are derived from infinite square well, which are sinusoidal functions and easier to deal with in software. In general, this approach can be extended to as many levels as we want. 

The Zeeman splitting of the system can be written as
\begin{equation}
    H_{Z}=-2 \mu_{B}(\kappa_1 \bm{J} \cdot \bm{B}+ \kappa_2 \bm{J}_3 \cdot \bm{B})
\end{equation}
If we extend our Luttinger-Kohn Hamiltonian into the 6-by-6 case. The matrix representation of $J$ and $J^3$ will be
\begin{equation}
    J_{x}=\frac{1}{2}\dsp{\left[\begin{array}{cccccc}
0 & 0 & \sqrt{3} & 0 & 0 & 0 \\
0 & 0 & 0 & \sqrt{3} & 0 & 0 \\
\sqrt{3} & 0 & 0 & 2 & 0 & 0 \\
0 & \sqrt{3} & 2 & 0 & 0 & 0 \\
0 & 0 & 0 & 0 & 0 & 1 \\
0 & 0 & 0 & 0 & 1 & 0
\end{array}\right]} \quad J_{y}=\frac{1}{2} i \dsp{\left[\begin{array}{cccccc}
0 & 0 & -\sqrt{3} & 0 & 0 & 0 \\
0 & 0 & 0 & \sqrt{3} & 0 & 0 \\
\sqrt{3} & 0 & 0 & -2 & 0 & 0 \\
0 & -\sqrt{3} & 2 & 0 & 0 & 0 \\
0 & 0 & 0 & 0 & 0 & -1 \\
0 & 0 & 0 & 0 & 1 & 0
\end{array}\right]} \quad J_{z}=\frac{1}{2}\left[\begin{array}{cccccc}
3 & 0 & 0 & 0 & 0 & 0 \\
0 & -3 & 0 & 0 & 0 & 0 \\
0 & 0 & 1 & 0 & 0 & 0 \\
0 & 0 & 0 & -1 & 0 & 0 \\
0 & 0 & 0 & 0 & 1 & 0 \\
0 & 0 & 0 & 0 & 0 & -1
\end{array}\right]
\end{equation} 
The size of the matrix which represent our total Hamiltonian $H$ will be the product of the band involved (4 heay-hole bands and 2 light-hole bands), the wave function we choose in x,y,z directions.
 
\section{Single phonon relaxation}
In this section, we discuss the relaxation time $T_1$ due to phonon-hole interactions. Consider the general case, a hole qubit interacts with a thermal bath of bulk acoustic phonons with energy $\hbar \omega_{\alpha, \bm{q}}$, where $\alpha$ indicates the polarization direction and $\bm{q}$ is a 3D wave-vector. 

The qubit and the phonons are coupled by a Hamiltonian $H^\alpha_{hp}$. The hole-phone Hamiltonian is derived similarly to the Bir-Pikus Hamiltonian that we used to describe the static strain during the fabrication process. Let us write down the Bir-Pikus Hamiltonian along a general polarization direction of phonons again:
\begin{equation}
    H^\alpha_{\text{BP}}\,=\,\mqty[ P^\alpha_\varepsilon+Q^\alpha_\varepsilon & 0 & -S^\alpha_\varepsilon & R^\alpha_\varepsilon & \dsp{-\frac{1}{\sqrt{2}}} S^\alpha_\varepsilon & \sqrt{2} R^\alpha_\varepsilon \\ 0 & P^\alpha_\varepsilon+Q^\alpha_\varepsilon & R^{\alpha*}_\varepsilon & S^{\alpha*}_\varepsilon & -\sqrt{2} R^{\alpha*}_\varepsilon & -\dsp{\frac{1}{\sqrt{2}}} S^{\alpha*}_\varepsilon \\ -S^{\alpha*}_\varepsilon & R^\alpha_\varepsilon & P^\alpha_\varepsilon-Q^\alpha_\varepsilon & 0 & -\sqrt{2}Q^\alpha_\varepsilon & \dsp{\sqrt{\frac{3}{2}}} S^\alpha_\varepsilon \\ R^{\alpha*}_\varepsilon & S^\alpha_\varepsilon & 0 & P^\alpha_\varepsilon-Q^\alpha_\varepsilon & \dsp{\sqrt{\frac{3}{2}}} S^{\alpha*}_\varepsilon & \sqrt{2}Q^\alpha_\varepsilon \\ -\dsp{\frac{1}{\sqrt{2}}} S^{\alpha*}_\varepsilon & -\sqrt{2}R^\alpha_\varepsilon & -\sqrt{2}Q^{\alpha*}_\varepsilon & \dsp{\sqrt{\frac{3}{2}}} S^\alpha_\varepsilon & P^\alpha_\varepsilon & 0 \\ \sqrt{2}R^{\alpha*}_\varepsilon & -\dsp{\frac{1}{\sqrt{2}}}S^\alpha_\varepsilon & \dsp{\sqrt{\frac{3}{2}}} S^{\alpha*}_\varepsilon & \sqrt{2}Q^{\alpha*}_\varepsilon & 0 & P^\alpha_\varepsilon]
\end{equation}
where
\begin{equation}
\begin{aligned}
    P^\alpha_\varepsilon =& -a_v \left(\varepsilon^\alpha_{xx}+\varepsilon^\alpha_{yy}+\varepsilon^\alpha_{zz}\right) \\
    Q^\alpha_\varepsilon =& -\dsp\frac{b_v}{2} \left(\varepsilon^\alpha_{xx}+\varepsilon^\alpha_{yy}-2 \varepsilon^\alpha_{zz}\right) \\
    R^\alpha_\varepsilon =& \dsp\frac{\sqrt{3}}{2} b_v \left(\varepsilon^\alpha_{xx}-\varepsilon^\alpha_{yy}\right) - i d_v \varepsilon^\alpha_{xy} \\
    S^\alpha_\varepsilon =& -d_v\left(\varepsilon^\alpha_{xz}-i \varepsilon^\alpha_{yz}\right)\\
    P^\alpha_\varepsilon+Q^\alpha_\varepsilon =& -\pqty{a_v+\dsp\frac{b_v}{2}} \varepsilon^\alpha_{xx} -\pqty{a_v+\frac{b_v}{2}} \varepsilon^\alpha_{yy} - (a_v-b_v) \varepsilon^\alpha_{zz} \\
    P^\alpha_\varepsilon-Q^\alpha_\varepsilon =& -\pqty{a_v-\dsp\frac{b_v}{2}} \varepsilon^\alpha_{xx} -\pqty{a_v-\frac{b_v}{2}} \varepsilon^\alpha_{yy} -\pqty{a_v+b_v} \varepsilon^\alpha_{zz}
\end{aligned} 
\end{equation}
To evaluate the strain (in the static case, it is just a number), we have to consider the local displacement field $\bm{u}(\va r)$, the relations between the local strain and the displacement field is given by
\begin{equation}
    \varepsilon^\alpha_{i,j}(\bm{r})\,=\,\frac{1}{2} \pqty{ \pdv{u_i(\bm{r})}{r_j} + \pdv{u_j(\bm{r})}{r_i} } \quad i,j \in (x,y,z)
\end{equation}
Here the position vector $r_x, r_y, r_z$ are really $x, y, z$. The displacement field will read
\begin{equation}
    \bm{u}\,=\,u_x \bm \hat{x} + u_y \bm \hat{y} + u_z \bm \hat{z}
\end{equation}
The local strain can be written as
\begin{equation}
    \varepsilon^\alpha_{i,j}\,=\,\frac{i}{2} \sqrt{\frac{\hbar}{2 V_c \rho v^\alpha q}} q \pqty{c_i^\alpha \frac{q_j}{q} + c_j^\alpha \frac{q_i}{q} } \pqty{ e^{-i \bm{q} \cdot \bm{r} } \hat{a}^{\dagger,\alpha}_{\va q} + e^{i \bm{q} \cdot \bm{r} } \hat{a}^{\alpha}_{\va q}} 
\end{equation}
where $\bm \hat c ^\alpha_{\bm{q}}\,=\,( \hat c_x, \hat c_y, \hat c_z)$ is the unit phonon polarization vector. $\rho$ is the density of the host material, $V_c$ is the crystal volume, $a^{\dagger\alpha}_{\bm{q}}$ is the phonon creation operators. The phonon frequency $\omega^\alpha_{\bm{q}}\,=\,v^\alpha q$ can be used to evaluate the energy of the phonon (the vibration mode). In silicon, we have
\begin{equation}
    v^\ell\,=\,9000 \text{m}/\text{s} \quad v^{t_1}\,=\,5400 \text{m}/\text{s} \quad v^{t_2}\,=\,5400 \text{m}/\text{s}
\end{equation}
We can further simplify the matrix by introducing the epsilon matrix, which are some numbers defined by the phonon polarization angles
\begin{equation}
    \epsilon_{i,j}^\alpha\,=\,\frac{1}{2} \pqty{ \hat c_i^\alpha \hat{q}_j + \hat c_j^\alpha \hat{q}_i }
\end{equation}
which can be written as
\begin{equation}
\epsilon^\alpha_{\va q}=\frac{1}{2}\left[\begin{array}{ccc}
2 {c}_x^\alpha \hat{q}_x & {c}_x^\alpha \hat{q}_y+{c}_y \hat{q}_x & {c}_x^\alpha \hat{q}_z+{c}^\alpha_z \hat{q}_x \\
{c}^\alpha_y \hat{q}_x+{c}^\alpha_x \hat{q}_y & 2 {c}^\alpha_y \hat{q}_y & {c}^\alpha_y \hat{q}_z+{c}^\alpha_z \hat{q}_y \\
{c}^\alpha_z \hat{q}_x+{c}^\alpha_x \hat{q}_z & {c}^\alpha_z \hat{q}_y+{c}_y \hat{q}_z & 2 {c}^\alpha_z \hat{q}_z
\end{array}\right]
\end{equation}
Now, consider the phonon vector take the general form $\bm \hat{q}\,=\,(\sin \theta \cos \varphi, \sin \theta \sin \varphi, \cos \theta)$ where $\theta$ is the azimuthal angle and $\phi$ is the in-plane polar angle. The longitudinal polarization is parallel to the phonon vector therefore we have
\begin{equation}
    \bm \hat c^\ell\,=\,q^{-1}\left(q_x, q_y, q_z\right)= (\sin \theta \cos \varphi, \sin \theta \sin \varphi, \cos \theta)
\end{equation}
The first transversal polarization vector is obtained by setting $\theta+\pi / 2, \varphi_{t_1}=\varphi$, which will give us
\begin{equation}
\begin{aligned}
\bm \hat{c}^{t_1} =& q^{-1}\left(q_x^2+q_y^2\right)^{-\frac{1}{2}}\left(q_x q_z, q_y q_z,-\left(q_x^2+q_y^2\right)\right)\,=\,( \cos(\theta)\cos(\varphi), \cos(\theta)\sin(\varphi), -\sin(\theta))   
\end{aligned}
\end{equation}
Another polarization vector can be obtained by $\left(\theta_{t_2}=\pi / 2, \varphi_{t_2}=\varphi+\pi / 2\right)$ 
\begin{equation}
\bm \hat{c}^{t_2}\,=\,\left(q_x^2+q_y^2\right)^{-\frac{1}{2}}\left(q_y,-q_x, 0\right)\,=\,(-\sin(\varphi), \cos(\varphi), 0 )
\end{equation}
In summary, we have
\begin{equation}
\begin{aligned}
    \bm \hat{c}^{\ell}=&q^{-1}\left(q_x, q_y, q_z\right)=(\sin \theta \cos \varphi, \sin \theta \sin \varphi, \cos \theta) \\ 
    \bm \hat{c}^{t_1}=&q^{-1}\left(q_x^2+q_y^2\right)^{-\frac{1}{2}}\left(q_x q_z, q_y q_z,-\left(q_x^2+q_y^2\right)\right)=(\cos (\theta) \cos (\varphi), \cos (\theta) \sin (\varphi),-\sin (\theta)) \\
    \bm \hat{c}^{t_2}=&\left(q_x^2+q_y^2\right)^{-\frac{1}{2}}\left(-q_y,q_x, 0\right)=(-\sin (\varphi), \cos (\varphi), 0)
\end{aligned}
\end{equation}
With all these polarization vectors, we can further calculate the $\epsilon$ matrix, in cartesian coordinates with $\bm \hat{q}\,=\,(\hat{q}_x, \hat{q}_y, \hat{q}_z)$ or $\bm \hat{q}\,=\,\pqty{ \dsp\frac{q_x}{q}, \dsp\frac{q_y}{q}, \dsp\frac{q_z}{q}}$, we have
\begin{equation}
    \epsilon^{\ell}(q,q_x,q_y,q_z)\,=\,\dsp\frac{1}{2} \dsp\mqty[ 2\dsp\frac{q^2_x}{q^2} & 2 \dsp\frac{q_x q_y}{q^2} & 2\dsp\frac{q_xq_z}{q^2} \\ 2\dsp\frac{q_yq_x}{q^2} & 2 \dsp\frac{q_y^2}{q^2} & 2\dsp\frac{q_yq_z}{q^2} \\ 2\dsp\frac{q_zq_x}{q^2} & 2\dsp\frac{q_zq_y}{q^2} & 2 \dsp\frac{q^2_z}{q^2} ]\,=\, \dsp\mqty[ \dsp\frac{q^2_x}{q^2} & \dsp\frac{q_x q_y}{q^2} & \dsp\frac{q_xq_z}{q^2} \\ \dsp\frac{q_yq_x}{q^2} & \dsp\frac{q_y^2}{q^2} & \dsp\frac{q_yq_z}{q^2} \\ \dsp\frac{q_zq_x}{q^2} & \dsp\frac{q_zq_y}{q^2} & \dsp\frac{q^2_z}{q^2} ]
\end{equation}
\begin{equation}
    \epsilon^{t_1}(q,q_x,q_y,q_z)\,=\,\dsp\frac{1}{2} \dsp\mqty[ 2\dsp\frac{q_x^2q_z}{q^2\sqrt{q_x^2+q_y^2}} & 2\dsp\frac{q_xq_yq_z}{q^2\sqrt{q_x^2+q_y^2}} & \dsp\frac{q_x q_z^2-q_xq_x^2-q_xq_y^2}{q^2\sqrt{q_x^2+q_y^2}} \\ 2\dsp\frac{q_xq_yq_z}{q^2\sqrt{q_x^2+q_y^2}} & 2\dsp\frac{q_y^2q_z}{q^2\sqrt{q_x^2+q_y^2}} & \dsp\frac{q_yq_z^2-q_yq_x^2-q_yq_y^2}{q^2\sqrt{q_x^2+q_y^2}} \\ \dsp\frac{q_x q_z^2-q_xq_x^2-q_xq_y^2}{q^2\sqrt{q_x^2+q_y^2}} & \dsp\frac{q_yq_z^2-q_yq_x^2-q_yq_y^2}{q^2\sqrt{q_x^2+q_y^2}} & -2\dsp\frac{q_z(q_x^2+q_y^2)}{q^2\sqrt{q_x^2+q_y^2}} ]
\end{equation}
\begin{equation}
    \epsilon^{t_2}(q,q_x,q_y,q_z)\,=\,\dsp\frac{1}{2} \dsp\mqty[ -2\dsp\frac{q_xq_y}{q\sqrt{q_x^2+q_y^2}} & \dsp\frac{q_x^2 - q_y^2}{q\sqrt{q_x^2+q_y^2}} & -\dsp\frac{q_yq_z}{q\sqrt{q_x^2+q_y^2}} \\ \dsp\frac{q_x^2 - q_y^2}{q\sqrt{q_x^2+q_y^2}} & 2\dsp\frac{q_xq_y}{q\sqrt{q_x^2+q_y^2}} & \dsp\frac{q_xq_z}{q\sqrt{q_x^2+q_y^2}} \\ -\dsp\frac{q_yq_z}{q\sqrt{q_x^2+q_y^2}} & \dsp\frac{q_xq_z}{q\sqrt{q_x^2+q_y^2}} & 0 ]
\end{equation}
Now we can decompose our hole-phonon Hamiltonian into the following form
\begin{equation}
    H^\alpha_{hp}\,=\,D_{1,1}^\alpha \varepsilon^\alpha_{1,1} + D_{1,2}^\alpha \varepsilon^\alpha_{1,2}+D_{1,3}^\alpha \varepsilon^\alpha_{1,3}+D_{2,2}^\alpha \varepsilon^\alpha_{2,2}+D_{2,3}^\alpha \varepsilon^\alpha_{2,3}+D_{3,3}^\alpha \varepsilon^\alpha_{3,3}
\end{equation}
where
\begin{equation}
    D_{11}\,=\,\mqty[ -\pqty{a_v+\dsp\frac{b_v}{2}} & 0 & 0 & \dsp\frac{\sqrt{3}}{2} b_v & 0& \sqrt{\dsp\frac{3}{2}}b_v \\ 0 & -\pqty{a_v+\dsp\frac{b_v}{2}} & \dsp\frac{\sqrt{3}}{2} b_v & 0 & -\sqrt{\dsp\frac{3}{2}}b_v & 0 \\ 0 & \dsp\frac{\sqrt{3}}{2} b_v & -\pqty{a_v-\dsp\frac{b_v}{2}} & 0 & \dsp\frac{b_v}{\sqrt{2}} & 0 \\ \dsp\frac{\sqrt{3}}{2} b_v & 0 & 0 & -\pqty{a_v-\dsp\frac{b_v}{2}} & 0 & -\dsp\frac{b_v}{\sqrt{2}} \\ 0 & -\sqrt{\dsp\frac{3}{2}}b_v & \dsp\frac{b_v}{\sqrt{2}} & 0 & -a_v & 0 \\ \sqrt{\dsp\frac{3}{2}}b_v & 0 & 0 & -\dsp\frac{b_v}{\sqrt{2}} & 0 & -a_v ]
\end{equation}
\begin{equation}
    D_{12}\,=\,\mqty[ 0 & 0 & 0 & -i d_v & 0 & -i \sqrt{2} d_v \\ 0 & 0 & i d_v & 0 & -i \sqrt{2} d_v & 0 \\ 0 & -i d_v & 0 & 0 & 0 & 0 \\ i d_v & 0 & 0 & 0 & 0 & 0 \\ 0 & i \sqrt{2} d_v & 0 & 0 & 0 & 0 \\ i \sqrt{2} d_v & 0 & 0 & 0 & 0 & 0 ]
\end{equation}
\begin{equation}
    D_{13}\,=\,\mqty[ 0 & 0 & d_v & 0 & \dsp\frac{d_v}{2} & 0 \\ 0 & 0 & 0 & -d_v & 0 & \dsp\frac{d_v}{\sqrt{2}} \\ d_v & 0 & 0 & 0 & 0 & -\sqrt{\dsp\frac{3}{2}}d_v \\ 0 & -d_v & 0 & 0 & -\sqrt{\dsp\frac{3}{2}} & 0 \\ \dsp\frac{d_v}{\sqrt{2}} & 0 & 0 & -\sqrt{\dsp\frac{3}{2}} & 0 & 0 \\ 0 & \dsp\frac{d_v}{\sqrt{2}} & -\sqrt{\dsp\frac{3}{2}} d_v & 0 & 0 & 0 ]
\end{equation}
\begin{equation}
    D_{22}\,=\,\mqty[ -\pqty{a_v+\dsp\frac{b_v}{2}} & 0 & 0 & -\dsp\frac{\sqrt{3}}{2} b_v & 0& -\sqrt{\dsp\frac{3}{2}}b_v \\ 0 & -\pqty{a_v+\dsp\frac{b_v}{2}} & -\dsp\frac{\sqrt{3}}{2} b_v & 0 & \sqrt{\dsp\frac{3}{2}}b_v & 0 \\ 0 & -\dsp\frac{\sqrt{3}}{2} b_v & -\pqty{a_v-\dsp\frac{b_v}{2}} & 0 & \dsp\frac{b_v}{\sqrt{2}} & 0 \\ -\dsp\frac{\sqrt{3}}{2} b_v & 0 & 0 & -\pqty{a_v-\dsp\frac{b_v}{2}} & 0 & -\dsp\frac{b_v}{\sqrt{2}} \\ 0 & \sqrt{\dsp\frac{3}{2}}b_v & \dsp\frac{b_v}{\sqrt{2}} & 0 & -a_v & 0 \\ -\sqrt{\dsp\frac{3}{2}}b_v & 0 & 0 & -\dsp\frac{b_v}{\sqrt{2}} & 0 & -a_v ]
\end{equation}
\begin{equation}
    D_{23}\,=\,\mqty[ 0 & 0 & -i d_v & 0 & -i\dsp\frac{d_v}{\sqrt{2}} & 0 \\ 0 & 0 & 0 & -i d_v & 0 & i\dsp\frac{d_v}{\sqrt{2}} \\ i d_v & 0 & 0 & 0 & 0 & i\sqrt{\dsp\frac{3}{2}} d_v \\ 0 & i d_v & 0 & 0 & -i\sqrt{\dsp\frac{3}{2}} d_v & 0 \\ i\dsp\frac{d_v}{\sqrt{2}} & 0 & 0 & i \sqrt{\dsp\frac{3}{2}} d_v & 0 & 0 \\ 0 & -i \dsp\frac{d_v}{\sqrt{2}} & -i \dsp\sqrt{\frac{3}{2}} d_v & 0 & 0 & 0 ]
\end{equation}
\begin{equation}
    D_{33}\,=\,\mqty[ -(a_v-b_v) & 0 & 0 & 0 & 0 & 0 \\ 0 & -(a_v-b_v) & 0 & 0 & 0 & 0 \\ 0 & 0 & -(a_v+b_v) & 0 & -\sqrt{2} b_v & 0 \\ 0 & 0 & 0 & -(a_v+b_v) & 0 & \sqrt{2} b_v \\ 0 & 0 & -\sqrt{2} b_v & 0 & -a_v & 0 \\ 0 & 0 & 0 &\sqrt{2} b_v & 0 & -a_v ]
\end{equation}
In this way, our Hamiltonian can be written as
\begin{equation}
    H_{hp}\,=\,\sum_{i,j} D_{i,j} \epsilon_{i,j} \times \bqty{ i q \sqrt{\frac{\hbar}{2 V_c \rho v^\alpha q}} \pqty{ e^{-i \bm{q} \cdot \bm{r} } \hat{a}^{\dagger,\alpha}_{\va q} + e^{i \bm{q} \cdot \bm{r} } \hat{a}^{\alpha}_{\va q}} } 
\end{equation}
Combined with the relaxation expression we used in the main text, we are able to calculate the relaxation time for all three directions.

\end{document}